\documentclass[]{tMPH2e}
\usepackage[usenames,dvipsnames]{xcolor}
\usepackage{graphics, inputenc}
\usepackage{graphicx,amsmath,amssymb,tabularx}
\usepackage{tikz}
\usepackage{multicol}
\usepackage{epstopdf}
\usepackage{float}
\usepackage{subfigure}
\usepackage{mathtools}
\definecolor{dgreen}{rgb}{0,.5,0}
\definecolor{dred}{rgb}{.7,.0,.0}

\newcommand{\manu}[1]{{\textcolor{dgreen}{ Manu: #1 }} }
\newcommand{\subi}[2]{#1_{\rm #2}}
\newcommand{\supi}[2]{#1^{\rm #2}}
\newcommand{\subsupi}[3]{#1_{\rm #2}^{\rm #3}}
\DeclarePairedDelimiter\bra{\langle}{\rvert}
\DeclarePairedDelimiter\ket{\lvert}{\rangle}
\DeclarePairedDelimiterX\braket[2]{\langle}{\rangle}{#1 \delimsize\vert #2}
\def\ddroit{{\rm d}}
\def\Tr{{\rm Tr}}

\DeclareMathOperator*{\minover}{min}

\begin{document}
\doi{doi}
 \issn{}
\issnp{}
\jvol{vol}
\jnum{num} \jyear{2015} \jmonth{September}

\markboth{B.~Senjean {\it et al.}}{Molecular Physics}

\articletype{Manuscript}

\title{{\itshape 
Combining linear interpolation with extrapolation methods in range-separated ensemble density-functional theory  
}
}

\author{
Bruno Senjean$^1$, Erik D. Hedeg{\aa}rd$^2$, Md. Mehboob Alam$^1$, Stefan
Knecht$^2$, and Emmanuel Fromager$^{1{\ast}}$\thanks{$^\ast$Corresponding author.
Email: fromagere@unistra.fr 
\vspace{6pt}}
\\\vspace{6pt}  
{\em{
$^1$Laboratoire de Chimie Quantique,
Institut de Chimie, CNRS / Universit\'{e} de Strasbourg,
4 rue Blaise Pascal, 67000 Strasbourg, France\\
\vspace{0.3cm}
$^2$Laboratory of Physical Chemistry,\\
ETH Z\"urich,\\
Vladimir-Prelog Weg 2, CH-8093 Z{\"u}rich, Switzerland
}}\\\vspace{6pt}  
}

\maketitle

\begin{abstract}

The combination of a recently proposed linear interpolation method
(LIM) [Senjean {\it et al.}, Phys. Rev. A {\bf 92},            
012518 (2015)], which enables the calculation of weight-independent excitation
energies in range-separated ensemble density-functional approximations,
with the extrapolation scheme of Savin [J. Chem.            
Phys. {\bf 140}, 18A509 (2014)] is presented in this work. It is shown that LIM excitation
energies vary quadratically with the inverse of the range-separation
parameter $\mu$ when the latter is large. As a result, the extrapolation
scheme, which
is usually applied to long-range interacting energies, can be adapted 
straightforwardly to LIM. This extrapolated LIM (ELIM) has been tested on a small test
set consisting of He, Be,
H$_2$ and HeH$^+$. Relatively accurate results have been obtained for the first
singlet excitation energies with the typical $\mu=0.4$ value. The improvement of
LIM after extrapolation is remarkable, in particular for the
doubly-excited $2^1\Sigma^+_g$ state in the stretched H$_2$ molecule.   
Three-state ensemble calculations in H$_2$ also show that ELIM does not
necessarily improves relative excitation energies, even though
individual excitation energies are more accurate after extrapolation.
Finally, an alternative decomposition of the short-range ensemble
exchange-correlation energy is proposed in order to correct for
ghost-interaction errors in multideterminant range-separated ensemble
density-functional theory calculations. The implementation and calibration of such a scheme is
currently in progress.

\bigskip

\begin{keywords}
Ensemble Density-Functional Theory,
Range Separation,
Linear Interpolation Method,
Ghost Interaction
\end{keywords}\bigskip

\end{abstract}

\section{Introduction}\label{sec:intro}

Nowadays, excitation energies are mostly computed by means of time-dependent density-functional theory (TDDFT)
\cite{marques2004time,Casida_tddft_review_2012}, essentially because of
its low computational cost and its relatively good accuracy. Despite
this success, standard TDDFT fails in describing multiple
excitations~\cite{maitra2004double} and, in general, multiconfigurational
effects~\cite{gritsenko2000excitation}. Ensemble
DFT~\cite{JPC79_Theophilou_equi-ensembles,PRA_GOK_RRprinc,
PRA_GOK_EKSDFT, GOK3} is an
alternative to TDDFT which can, in principle, provide
exact excitation energies in a time-independent framework. Despite the fact
that developing exchange-correlation functionals for ensembles is
challenging~\cite{Nagy_enseXpot, ParagiXens,
ParagiXCens,pernal2015excitation}, thus explaining why ensemble DFT is
not yet standard,
the use of ensembles in DFT for excited states has reappeared recently
in the
literature (see, for example,
Refs.~\cite{franck2014generalised,Burke_ensemble,yang2014exact,JCP14_Filatov_conical_inter_REKS,pastorczak2014ensemble,filatov2015ensemble,Filatov-2015-Wiley}).
One interesting feature of ensemble DFT is that
it offers, when combined with range
separation~\cite{savinbook}, a rigorous
framework for developing state-averaged multi-determinant DFT methods~\cite{PRA13_Pernal_srEDFT,franck2014generalised,senjean2015linear}.\\
In practical range-separated ensemble DFT calculations, ground-state
functionals are used, simply because short-range ensemble
exchange-correlation density-functional approximations have not
been developed so far. This leads to curved range-separated ensemble
energies~\cite{senjean2015linear} and, consequently, to ensemble-weight-dependent excitation
energies. Some of the authors recently proposed a linear interpolation
method (LIM) \cite{senjean2015linear} which, by construction, provides
linear ensemble energies and, therefore, well-defined approximate
excitation energies. Promising results have been obtained with LIM, especially
for the $2^1\Sigma^+$ charge-transfer state in the stretched 
HeH$^+$ molecule and the doubly-excited $2^1\Sigma_g^+$ state in the
stretched H$_2$ molecule. However, in the latter case, the excitation energy was
still significantly underestimated \cite{senjean2015linear}.\\
In this work, we propose to adapt the
extrapolation scheme of Savin
\cite{savin2014towards,RebTouTeaHelSav-JCP-14,rebolini2015calculating} to ensembles. The
basic idea is to expand the approximate LIM ensemble energy in $1/\mu$,
where $\mu$ is the parameter that controls the range separation, and to
use this expansion for improving the convergence towards the exact
result (i.e.~the $\mu\rightarrow+\infty$ limit).     
The paper is organized as follows: After a brief introduction to
ground-state range-separated DFT in Sec.~\ref{subsec:Range-separation},
ensemble DFT is presented in Sec.~\ref{subsec:GOKDFT} and its
range-separated extension is discussed in Sec.~\ref{subsec:srGOKDFT}.
LIM is then introduced in Sec.~\ref{subsec:LIM} and its combination with
Savin's extrapolation technique is presented in Sec.~\ref{subsec:LIM+Extrapolation}. After a summary in Sec.~\ref{Summary} and the computational
details in Sec.~\ref{sec:computation}, results obtained for He, Be,
H$_2$ and HeH$^+$ are discussed in Sec.~\ref{sec:Results}. As a
perspective, a correction scheme 
for ghost-interaction errors in
multideterminant range-separated ensemble DFT
is proposed in
Sec.~\ref{sec:Perspective}. Finally, conclusions are given in Sec.~\ref{sec:Conclusion}.

\section{Theory}

\subsection{Range-separated density-functional theory for the ground state}\label{subsec:Range-separation}

The exact ground-state energy of an electronic system can be obtained
variationally as follows, according to the Hohenberg--Kohn (HK) theorem \cite{hktheo},
\begin{eqnarray}\label{eq:gs_energy}
E_0 = \minover\limits_{n} \left\lbrace F[n] + \int \ddroit \mathbf{r} \phantom{i} \subi{v}{ne} (\mathbf{r})n(\mathbf{r}) \right \rbrace ,
\end{eqnarray}
where $\subi{v}{ne}(\mathbf{r})$ is the nuclear potential,
$n(\mathbf{r})$ is a trial electron density and $F[n]$ is the universal
Levy--Lieb (LL) functional defined by
\begin{eqnarray}\label{eq:LL_functional}
F[n] = \minover\limits_{\Psi \rightarrow n} \bra{\Psi} \hat{T} + \subi{\hat{W}}{ee} \ket{\Psi} ,
\end{eqnarray}
where $\hat{T}$ is the kinetic energy operator and $\subi{\hat{W}}{ee}$
is the two-electron repulsion operator. In Kohn--Sham DFT~\cite{KS}
(KS-DFT), we
consider a noninteracting system which has the same density as the
physical one, and the wavefunction is replaced by a Slater determinant
$\Phi^{\rm KS}[n]$.  The universal LL functional becomes
\begin{eqnarray}\label{eq:LL_functional_Phi}
F[n] = 
\bra{\Phi^{\rm KS}[n]} \hat{T}\ket{\Phi^{\rm KS}[n]}
+ \subi{E}{Hxc}[n] = T_s [n] + \subi{E}{Hxc}[n],
\end{eqnarray}
where $\subi{E}{Hxc}[n]$ is the universal Hartree-exchange-correlation density functional and $T_s[n]$ is the noninteracting kinetic energy. 
An exact multideterminantal extension of KS-DFT can be obtained by
decomposing the two-electron repulsion into long-range and short-range
parts, as proposed by Savin~\cite{savinbook}. This decomposition is
controlled by a parameter $\mu$,
\begin{eqnarray}\label{eq:w_ee_savin}
\subi{w}{ee}(r_{12}) & = & \dfrac{1}{r_{12}} = \subsupi{w}{ee}{lr,\mu}(r_{12}) + \subsupi{w}{ee}{sr,\mu}(r_{12}),\nonumber \\
\subsupi{w}{ee}{lr,\mu}(r_{12}) & = & {\rm erf}(\mu r_{12})/r_{12} = \frac{\displaystyle 2}{\displaystyle
 r_{12}\sqrt{\pi} }{\displaystyle\int^{\mu r_{12}}_0 e^{-t^{2}}\ddroit t} 
,
\end{eqnarray}
where ${\rm erf}$ is the error function and $\mu \in [0, +\infty [$. The
key idea in
range-separated DFT is to treat the long-range interaction explicitly
and to describe the short-range counterpart implicitly by means of a density functional, thus leading to the following expression for
the universal LL functional,
\begin{eqnarray}\label{eq:LL_functional_savin}
F[n] = \supi{F}{lr, \mu}[n] + \subsupi{E}{Hxc}{sr,\mu}[n],
\end{eqnarray} 
where
\begin{eqnarray}\label{eq:LL_functional_lr}
\supi{F}{lr, \mu}[n] &=& \minover\limits_{\Psi \rightarrow n} \bra{\Psi} \hat{T} + \subsupi{\hat{W}}{ee}{lr,\mu} \ket{\Psi}
\nonumber\\
&=&\bra{\Psi^\mu[n]} \hat{T} + \subsupi{\hat{W}}{ee}{lr,\mu}
\ket{\Psi^\mu[n]}
.
\end{eqnarray}
In analogy with KS-DFT, the short-range density functional can be decomposed 
as follows,
\begin{eqnarray}
\subsupi{E}{Hxc}{sr, \mu} [n] = \subsupi{E}{H}{sr, \mu} [n] + \subsupi{E}{x}{sr, \mu} [n] + \subsupi{E}{c}{sr, \mu} [n],
\end{eqnarray}
where the short-range Hartree term is given by
\begin{eqnarray}\label{eq:hartree_part}
\subsupi{E}{H}{sr,\mu}[n] = \dfrac{1}{2} \int \int \ddroit \mathbf{r}
\ddroit \mathbf{r'} n(\mathbf{r})n(\mathbf{r'}) \subsupi{w}{ee}{sr, \mu}
\left(\vert\mathbf{r} - \mathbf{r'} \vert \right) ,
\end{eqnarray}
the short-range exchange part is defined as
\begin{eqnarray}
\subsupi{E}{x}{sr, \mu} [n] = \bra{\Phi^{\rm KS} [n]} \subsupi{\hat{W}}{ee}{sr, \mu} \ket{\Phi^{\rm KS}[n]} - \subsupi{E}{H}{sr,\mu}[n],
\end{eqnarray}
and the remaining short-range correlation part can be connected with the
conventional (full-range) correlation functional as
follows~\cite{srDFT,manusroep2013},  
\begin{eqnarray}\label{eq:srcorr_conv_KS}
\subsupi{E}{c}{sr, \mu} [n] = E_{\rm c} [n] + \bra{\Phi^{\rm KS} [n]} \hat{T} + \subsupi{\hat{W}}{ee}{lr, \mu} \ket{\Phi^{\rm KS}[n]} -  \bra{\Psi^{\mu} [n]} \hat{T} + \subsupi{\hat{W}}{ee}{lr, \mu} \ket{\Psi^{\mu}[n]}.
\end{eqnarray}
From the decomposition in Equation~(\ref{eq:LL_functional_savin}), we
obtain the following expression for the exact ground-state energy,
\begin{eqnarray}\label{eq:gs_energy_savin}
E_0 & = & \minover\limits_{n} \left \lbrace \supi{F}{lr, \mu}[n] + \subsupi{E}{Hxc}{sr,\mu}[n] + \int \ddroit \mathbf{r} \phantom{i} \subi{v}{ne}(\mathbf{r}) n(\mathbf{r}) \right\rbrace , \nonumber\\
& = & \minover\limits_{\Psi} \left\lbrace \bra{\Psi} \hat{T} + \subsupi{\hat{W}}{ee}{lr,\mu} + \subi{\hat{V}}{ne} \ket{\Psi} + \subsupi{E}{Hxc}{sr,\mu}[n_{\Psi}] \right\rbrace  , \nonumber \\
& = & \bra{\subsupi{\Psi}{0}{\mu}} \hat{T} + \subsupi{\hat{W}}{ee}{lr,\mu} + \subi{\hat{V}}{ne} \ket{\subsupi{\Psi}{0}{\mu}} + \subsupi{E}{Hxc}{sr,\mu}[n_{\subsupi{\Psi}{0}{\mu}}] ,
\end{eqnarray}
where $\Psi_0^{\mu}$ is the exact minimizing wavefunction which has the
same density as the fully-interacting ground-state wavefunction $\Psi_0$, and $\hat{V}_{\rm ne} = \int \ddroit \mathbf{r} \phantom{i} \subi{v}{ne}(\mathbf{r}) \hat{n}(\mathbf{r})$ where $\hat{n}(\mathbf{r})$ is the density operator. 
It fulfills the following self-consistent equation,
\begin{eqnarray}\label{eq:selfconsistent_eq}
\hat{H}^{\mu} [n_{\Psi_0^\mu}]\vert \Psi_0^\mu\rangle =
\mathcal{E}_0^\mu \vert\Psi_0^\mu\rangle ,
\end{eqnarray}
where 
\begin{eqnarray}\label{eq:hamiltonian_lr}
\hat{H}^{\mu} [n_{\Psi_0^\mu}] = \hat{T} + \subsupi{\hat{W}}{ee}{lr,\mu}
+ \subi{\hat{V}}{ne} + \int \ddroit \mathbf{r} \dfrac{\delta
\subsupi{E}{Hxc}{sr, \mu} [n_{\Psi_0^\mu}]}{\delta n(\mathbf{r})} \hat{n}(\mathbf{r}) .
\end{eqnarray}
Range separation leads to an exact combination of wavefunction theory
and KS-DFT. The former is recovered when $\mu\rightarrow+\infty$, while
the latter is recovered when $\mu=0$. Note that the spectrum
$\left\{\mathcal{E}_k^\mu\right\}_{k=0,1,2,\ldots}$ of the auxiliary
long-range interacting Hamiltonian $\hat{H}^{\mu} [n_{\Psi_0^\mu}]$ can
be used as a starting point for approaching the exact physical
spectrum. This can be achieved within the time-dependent linear response
regime~\cite{fromager2013}. In this paper, we
will focus on time-independent approaches. One of them, which was initially proposed by
Savin~\cite{savin2014towards}, consists in using the expansion of the auxiliary
energies for large $\mu$ values,
\begin{eqnarray}\label{extrapol_auxiener_exact_GSdens}
 \mathcal{E}_k^{\mu}=E_k-\dfrac{\mu}{2}
\dfrac{\partial {\mathcal{E}}_k^{\mu}}{\partial \mu}
+\mathcal{O}\left(\dfrac{1}{\mu^3}\right),
\end{eqnarray}
where $E_k$ is the exact $k$th physical energy, thus leading to the
extrapolated energy through second order in
$1/\mu$~\cite{savin2014towards,RebTouTeaHelSav-JCP-14,rebolini2015calculating},
\begin{eqnarray}\label{extrapol_auxiener_exact_GSdens2}
 \mathcal{E}_{E,k}^{\mu}=\mathcal{E}_k^{\mu}+\dfrac{\mu}{2}
\dfrac{\partial {\mathcal{E}}_k^{\mu}}{\partial \mu}.
\end{eqnarray}
As discussed in the following, such an extrapolation scheme can also be
applied to range-separated ensemble energies in order to obtain more
accurate excitation energies.

\subsection{Ensemble density-functional theory}\label{subsec:GOKDFT}

In contrast to KS-DFT, ensemble DFT allows for the calculation of
excitation energies. This is achieved by assigning weights to ground and
excited states so that the weighted sum of energies and densities,
referred to as ensemble energy and ensemble density, respectively, can
be constructed. Let us consider an ensemble consisting of $M$ states
with ensemble weights $\mathbf{w}$ $\equiv (w_0, w_1, ..., w_{M-1})$
ordered as $w_0 \geq w_1 \geq ... \geq w_{M-1}$.
Note that Boltzmann weights can be used~\cite{PRA13_Pernal_srEDFT} but it is not
compulsory. The summation constraint $\sum_{k=0}^{M-1} w_k = 1$ will be
used in the following so that the ensemble density integrates to the
number $N$ of electrons. According to the Gross--Oliveira--Kohn (GOK) variational principle
\cite{PRA_GOK_RRprinc},
\begin{eqnarray}\label{eq:ensemble_energy_trace}
E^\mathbf{w} \leq \Tr \left[ \hat{\gamma}^\mathbf{w}\hat{H} \right],
\end{eqnarray}
where Tr denotes the trace, $\hat{H} = \hat{T} + \hat{W}_{\rm ee} +
\hat{V}_{\rm ne}$ is the physical Hamiltonian, and
$\hat{\gamma}^\mathbf{w}$ is a trial ensemble density matrix constructed
from a set of $M$ orthonormal trial wavefunctions $\{ \overline{\Psi}_k
\}_{0 \leq k \leq M-1}$:
\begin{eqnarray}\label{eq:density_matrix}
\hat{\gamma}^\mathbf{w} = \displaystyle \sum_{k = 0}^{M-1} w_k \ket{\overline{\Psi}_k}\bra{\overline{\Psi}_k}.
\end{eqnarray}
The exact ensemble energy is the lower bound in Equation~(\ref{eq:ensemble_energy_trace}), 
\begin{eqnarray}\label{eq:ensemble_energy_sumofallstates}
E^\mathbf{w} = \Tr \left[ \hat{\Gamma}^\mathbf{w}\hat{H} \right] = \sum_{k=0}^{M-1}w_kE_k ,
\end{eqnarray}
where $\hat{\Gamma}^\mathbf{w}= \sum_{k = 0}^{M-1} w_k \ket{\Psi_k}\bra{\Psi_k}$ is the exact physical ensemble density
matrix, $\Psi_k$ is the exact $k$th eigenfunction of $\hat{H}$ and $E_0
\leq E_1 \leq ... \leq E_{M-1}$. The GOK variational principle extends the HK theorem to ensembles~\cite{PRA_GOK_EKSDFT}, so that the
ensemble energy can be obtained variationally as follows,
\begin{eqnarray}\label{eq:ensemble_energy_withLLensemblefunctional}
E^{\mathbf{w}} = \minover\limits_{n} \left \lbrace F^\mathbf{w}[n] + \int \ddroit \mathbf{r} \phantom{i} \subi{v}{ne}(\mathbf{r}) n(\mathbf{r}) \right \rbrace ,
\end{eqnarray}
where the universal LL ensemble functional equals 
\begin{eqnarray}\label{eq:LLensemble_functional}
F^\mathbf{w}[n] = \minover\limits_{\hat{\gamma}^\mathbf{w} \rightarrow n} \left \lbrace \Tr \left[ \hat{\gamma}^\mathbf{w} \left( \hat{T} + \hat{W}_{\rm ee} \right) \right] \right \rbrace .
\end{eqnarray}
Note that, in Equation~(\ref{eq:LLensemble_functional}), the minimization is restricted to ensemble density matrices with ensemble density $n$:
\begin{eqnarray}\label{eq:ensemble_densities}
\Tr \left[\hat{\gamma}^\mathbf{w} \hat{n}(\mathbf{r}) \right] = n_{\hat{\gamma}^\mathbf{w}}(\mathbf{r}) = n(\mathbf{r}).
\end{eqnarray}
Note also that the minimizing density in
Equation~(\ref{eq:ensemble_energy_withLLensemblefunctional}) is the
exact ensemble density $n_{\hat{\Gamma}^{\mathbf{w}}} (\mathbf{r}) =
\sum_{k=0}^{M-1} w_k n_{\Psi_k} (\mathbf{r})$. By analogy with KS-DFT,
Gross~{\it et al.}~\cite{PRA_GOK_EKSDFT} considered the following partitioning of the LL ensemble functional,
\begin{eqnarray}\label{eq:LL_ensemble_KS}
F^\mathbf{w}[n] = T^\mathbf{w}_s [n] + \subsupi{E}{Hxc}{\mathbf{w}} [n] , 
\end{eqnarray}
where $T^\mathbf{w}_s [n]$ is the noninteracting ensemble kinetic energy,
\begin{eqnarray}\label{eq:kinetic_KS_ensemble}
T^\mathbf{w}_s [n] &=& \minover\limits_{\hat{\gamma}^\mathbf{w} \rightarrow n} \left \lbrace \Tr \left[\hat{\gamma}^\mathbf{w} \hat{T} \right] \right \rbrace 
\nonumber\\
&=&\Tr \left[\hat{\Gamma}_s^\mathbf{w}[n] \hat{T} \right]
,
\end{eqnarray}
$\hat{\Gamma}_s^\mathbf{w}[n]$ denotes the noninteracting ensemble density matrix with density $n$ and $\subsupi{E}{Hxc}{\mathbf{w}}[n]$ is the weights-dependent
Hartree-exchange-correlation ensemble functional. The conventional
(weights-independent) ground-state
Hartree functional is usually employed~\cite{PRA_GOK_EKSDFT}, thus
leading to the following decomposition,  
\begin{eqnarray}\label{eq:Hxcfun_gok_decomp}
\subsupi{E}{Hxc}{\mathbf{w}} [n] = \subsupi{E}{H}{} [n] + \subsupi{E}{x}{\mathbf{w}} [n] + \subsupi{E}{c}{\mathbf{w}} [n],
\end{eqnarray}
where the exact ensemble exchange density-functional energy is defined in terms of the
noninteracting ensemble density matrix,
\begin{eqnarray}\label{eq:exchange_fun_ensemble_GOK}
\subsupi{E}{x}{\mathbf{w}} [n]= \Tr \left[ \hat{\Gamma}_s^\mathbf{w} [n] \subsupi{\hat{W}}{ee}{} \right] - \subsupi{E}{H}{} [n].
\end{eqnarray}
Note that the Hartree term will always contain so-called
ghost-interaction errors~\cite{ensemble_ghost_interaction} when computed
with an ensemble density,
simply because it is quadratic in the input density $n$. The exact
ensemble exchange functional removes such errors, as readily seen in
Equation~(\ref{eq:exchange_fun_ensemble_GOK}). Since, in practice, approximate
functionals are used, ghost interactions might be significant, thus
requiring correction
schemes~\cite{ensemble_ghost_interaction,pastorczak2014ensemble,Nagy_functional}.
This will be discussed further in Sec.~\ref{sec:Perspective}, in the context of range-separated ensemble DFT.\\    

By using the KS partioning in Equation~(\ref{eq:LL_ensemble_KS}), the exact ensemble energy becomes
\begin{eqnarray}\label{eq:ensemble_energy_KS}
E^\mathbf{w} = \minover\limits_{n} \left\lbrace T^\mathbf{w}_s [n] + \int \ddroit \mathbf{r} \phantom{i} \subi{v}{ne} (\mathbf{r}) n(\mathbf{r}) + \subsupi{E}{Hxc}{\mathbf{w}} [n] \right \rbrace ,
\end{eqnarray}
or, equivalently,
\begin{eqnarray}\label{eq:ensemble_energy_KS_densitymatrix}
E^\mathbf{w} = \minover\limits_{\hat{\gamma}^\mathbf{w}} \left \lbrace \Tr \left[\hat{\gamma}^\mathbf{w} \left( \hat{T} + \subi{\hat{V}}{ne} \right) \right] + \subsupi{E}{Hxc}{\mathbf{w}} [n_{\hat{\gamma}^\mathbf{w}}] \right\rbrace .
\end{eqnarray}
The minimizing noninteracting GOK ensemble density matrix, 
\begin{eqnarray}\label{eq:density_matrix_noninteracting}
\hat{\Gamma}^\mathbf{w}_s = \displaystyle \sum_{k =0}^{M-1} w_k \ket{\Phi_k^\mathbf{w}} \bra{\Phi_k^\mathbf{w}} ,
\end{eqnarray}
reproduces the exact physical ensemble density:
$n_{\hat{\Gamma}^\mathbf{w}_s}(\mathbf{r}) =
n_{\hat{\Gamma}^{\mathbf{w}}}(\mathbf{r})$. Finally, by considering the
Lagrangian~\cite{senjean2015linear}
\begin{eqnarray}\label{eq:lagrangien_ensembleexacttheory}
\mathcal{L}^\mathbf{w} [\hat{\gamma}^\mathbf{w}] = \Tr \left[ \hat{\gamma}^\mathbf{w} \left(\hat{T} + \hat{V}_{\rm ne} \right) \right] + \subsupi{E}{Hxc}{\mathbf{w}}[n_{\hat{\gamma}^\mathbf{w}}] + \displaystyle \sum_{k=0}^{M-1} w_k \mathcal{E}_k^\mathbf{w} \left( 1 - \braket{\overline{\Psi}_k}{\overline{\Psi}_k} \right) , 
\end{eqnarray}
where $\mathcal{E}_k^\mathbf{w}$ are Lagrange multipliers associated
with the normalization of the trial wavefunctions $\overline{\Psi}_k$
from which the ensemble density matrix is built, we obtain from the stationarity
condition $\delta \mathcal{L}^\mathbf{w} [\hat{\Gamma}^\mathbf{w}_s] =
0$ the self-consistent GOK-DFT equations \cite{PRA_GOK_EKSDFT}:
\begin{eqnarray}\label{eq:self_consistent_exactensembletheory}
\left( \hat{T} + \subi{\hat{V}}{ne} + \int \ddroit \mathbf{r} \dfrac{\delta \subsupi{E}{Hxc}{\mathbf{w}}[n_{\hat{\Gamma}^\mathbf{w}_s}] }{\delta n(\mathbf{r})}\hat{n}(\mathbf{r}) \right) \ket{\Phi^\mathbf{w}_k} = \mathcal{E}_k^\mathbf{w} \ket{\Phi_k^{\mathbf{w}}}.
\end{eqnarray}

\subsection{Range-separated ensemble density-functional theory}\label{subsec:srGOKDFT}

In analogy with ground-state range-separated DFT, range separation can
be introduced into the LL ensemble functional~\cite{PRA13_Pernal_srEDFT,franck2014generalised}, thus leading to
the following decomposition,
\begin{eqnarray}\label{eq:LL_ensemble_savin}
F^{\mathbf{w}}[n] = \supi{F}{lr, \mu , \mathbf{w}} [n] + \subsupi{E}{Hxc}{sr, \mu , \mathbf{w}} [n] ,
\end{eqnarray}
where
\begin{eqnarray}\label{eq:LL_lr_ensemble_savin}
 \supi{F}{lr, \mu , \mathbf{w}} [n] &=&
\minover\limits_{\hat{\gamma}^\mathbf{w} \rightarrow n} \left\lbrace 
\Tr
\left[ \hat{\gamma}^\mathbf{w} \left( \hat{T} +
\subsupi{\hat{W}}{ee}{lr, \mu} \right) \right] 
\right\rbrace
\nonumber\\ 
&=&
\Tr
\left[ \hat{\Gamma}^{\mu,\mathbf{w}}[n] \left( \hat{T} +
\subsupi{\hat{W}}{ee}{lr, \mu} \right) \right] 
.
\end{eqnarray}
The short-range ensemble density functional is both $\mu$- and
${\bf w}$-dependent. In analogy with GOK-DFT~\cite{PRA_GOK_EKSDFT}, it can be partitioned as
follows,
\begin{eqnarray}\label{eq:EHxc_decomposition_ensemble}
\subsupi{E}{Hxc}{sr, \mu , \mathbf{w}} [n] = \subsupi{E}{H}{sr, \mu} [n] + \subsupi{E}{x}{sr, \mu , \mathbf{w}} [n] + \subsupi{E}{c}{sr, \mu , \mathbf{w}} [n],
\end{eqnarray}
where the ground-state ($\mathbf{w}$-independent) short-range Hartree
functional defined in Equation~(\ref{eq:hartree_part}) is used. The
exact short-range exchange ensemble energy can be
expressed in terms of the noninteracting ensemble density matrix (see
Equation~(\ref{eq:kinetic_KS_ensemble})) as
\begin{eqnarray}\label{eq:exc_functional_ensemble}
\subsupi{E}{x}{sr, \mu , \mathbf{w}} [n]  = \Tr \left[ \hat{\Gamma}_s^\mathbf{w} [n] \subsupi{\hat{W}}{ee}{sr, \mu} \right] - \subsupi{E}{H}{sr, \mu} [n] ,
\end{eqnarray}
and, according to Equations~(\ref{eq:LL_ensemble_KS}),
(\ref{eq:Hxcfun_gok_decomp}),
(\ref{eq:exchange_fun_ensemble_GOK}), (\ref{eq:LL_ensemble_savin}) and
(\ref{eq:LL_lr_ensemble_savin}), the complementary
short-range correlation energy for the ensemble can be related with the
conventional (full-range) correlation energy as follows,
\begin{eqnarray}\label{eq:corr_functional_ensemble}
\subsupi{E}{c}{sr, \mu , \mathbf{w}} [n] = \subsupi{E}{c}{\mathbf{w}}[n]
+ \Tr \left[ \hat{\Gamma}_s^\mathbf{w} [n] \left( \hat{T} +
\subsupi{\hat{W}}{ee}{lr,\mu} \right) \right] - \Tr \left[
\hat{\Gamma}^{\mu,\mathbf{w}} [n] \left( \hat{T} + \subsupi{\hat{W}}{ee}{lr, \mu} \right) \right] .
\end{eqnarray}
Note that Equation~(\ref{eq:srcorr_conv_KS}) is recovered when the
ensemble reduces to the ground state ($w_0=1$). Note also that, as in GOK-DFT, the exact short-range ensemble energy
removes the ghost-interaction errors introduced in the short-range
Hartree energy (see Equations~(\ref{eq:exchange_fun_ensemble_GOK}) and
(\ref{eq:exc_functional_ensemble})). In practice~\cite{PRA13_Pernal_srEDFT,senjean2015linear}, the use of (semi-) local
ground-state short-range exchange density functionals cannot, in
principle, remove such errors. In Sec.~\ref{sec:Perspective}, an alternative
decomposition of the short-range ensemble exchange-correlation energy
will be proposed and discussed. The latter provides a rigorous framework
for performing 
ghost-interaction-free multi-determinant range-separated ensemble DFT
calculations. Work is currently in progress in this direction.\\ 

Returning to the range-separated LL ensemble functional expression in
Equation~(\ref{eq:LL_ensemble_savin}), we finally obtain from
Equation~(\ref{eq:ensemble_energy_withLLensemblefunctional}) the
following exact expression for the ensemble energy,
\begin{eqnarray}\label{eq:ensemble_energy_savin_ensemble}
E^\mathbf{w} & = & \minover\limits_{\hat{\gamma}^\mathbf{w}} \left\lbrace \Tr \left[ \hat{\gamma}^\mathbf{w} \left( \hat{T} + \subsupi{\hat{W}}{ee}{lr, \mu} + \subi{\hat{V}}{ne}\right) \right] +  \subsupi{E}{Hxc}{sr, \mu , \mathbf{w}} [n_{\hat{\gamma}^\mathbf{w}}] \right\rbrace  \nonumber \\
& = & \Tr \left[ \hat{\Gamma}^{\mu, \mathbf{w}} \left( \hat{T} +
\subsupi{\hat{W}}{ee}{lr, \mu} + \subi{\hat{V}}{ne}\right) \right] +
\subsupi{E}{Hxc}{sr, \mu , \mathbf{w}} [n_{\hat{\Gamma}^{\mu,
\mathbf{w}}}], 
\end{eqnarray}
where the minimizing long-range-interacting ensemble density matrix
$\hat{\Gamma}^{\mu, \mathbf{w}} = \sum_{k = 0}^{M-1} w_k \ket{\Psi^{\mu
, \mathbf{w}}_k}\bra{\Psi^{\mu , \mathbf{w}}_k}$ reproduces the exact
physical ensemble density: $n_{\hat{\Gamma}^{\mu,
\mathbf{w}}}(\mathbf{r}) = n_{\hat{\Gamma}^{\mathbf{w}}}(\mathbf{r})$.
This density matrix fulfills the following self-consistent equation, in analogy with Equation~(\ref{eq:self_consistent_exactensembletheory}),
\begin{eqnarray}\label{eq:self_consistent_savin_ensemble}
\left( \hat{T} + \subsupi{\hat{W}}{ee}{lr, \mu} + \subi{\hat{V}}{ne} + \int \ddroit \mathbf{r} \dfrac{\delta \subsupi{E}{Hxc}{sr, \mu , \mathbf{w}}[n_{\hat{\Gamma}^{\mu, \mathbf{w}}}] }{\delta n(\mathbf{r})}\hat{n}(\mathbf{r}) \right) \ket{\Psi^{\mu , \mathbf{w}}_k} = \mathcal{E}_k^{\mu , \mathbf{w}} \ket{\Psi_k^{\mu ,\mathbf{w}}}.
\end{eqnarray}
When $\mu \rightarrow +\infty$, we recover the Schr\"odinger equation
and, when $\mu = 0$, GOK-DFT equations are obtained.\\

Let us now consider the particular case of two non-degenerate states.
The ensemble weights are then reduced to two weights, $w_0$ for the
ground state and $w_1$ for the first excited state. Both can be
expressed in terms of a single weight $w$, 
\begin{eqnarray}\label{eq:weights_two}
w_0 = 1 - w, \phantom{e} w_1 = w,
\end{eqnarray}
with $w \in [0, 1/2]$. The exact ensemble energy is a linear function of $w$,
\begin{eqnarray}\label{eq:ensemble_energy}
E^w = (1-w)E_0 + wE_1,
\end{eqnarray}
and the excitation energy $\omega$ is simply the first derivative of the ensemble energy with respect to $w$:
\begin{eqnarray}\label{eq:excitation_energy_exact}
\omega & = & \dfrac{\ddroit E^w}{\ddroit w} , \nonumber \\
& = & E_1 - E_0.
\end{eqnarray}
Linearity in $w$ enables also to obtain the excitation energy by linear
interpolation between ground-state and equiensemble energies, 
\begin{eqnarray}\label{eq:excitation_energy_exact_ensemble}
\omega & = & 2(E^{w = 1/2} - E^{w = 0}), \nonumber \\
 & = & 2(E^{w = 1/2} - E_0).
\end{eqnarray}
In this particular case of a two-state ensemble, the exact range-separated
ensemble energy in Equation~(\ref{eq:ensemble_energy_savin_ensemble})
can be rewritten as
\begin{eqnarray}\label{eq:ensemble_energy_savin}
E^w & = & (1 - w) \bra{\Psi_0^{\mu , w}} \hat{T} + \subsupi{\hat{W}}{ee}{lr,\mu} + \subi{\hat{V}}{ne} \ket{\Psi_0^{\mu , w}} \nonumber\\
& & + w \bra{\Psi_1^{\mu , w}} \hat{T} + \subsupi{\hat{W}}{ee}{lr,\mu} +
\subi{\hat{V}}{ne} \ket{\Psi_1^{\mu , w}} + \subsupi{E}{Hxc}{sr,\mu ,
{\it w}}[n_{\hat{\Gamma}^{\mu,w}}] ,
\end{eqnarray}
or, equivalently (see Equation~(\ref{eq:self_consistent_savin_ensemble})),
\begin{eqnarray}\label{eq:ensemble_energy_aux}
E^w = (1-w) \mathcal{E}_0^{\mu , w} + w \mathcal{E}_1^{\mu , w} - \int
\ddroit\mathbf{r} \dfrac{\delta \subsupi{E}{Hxc}{sr,\mu , {\it
w}}[n_{\hat{\Gamma}^{\mu,w}}]}{\delta
n(\mathbf{r})}n_{\hat{\Gamma}^{\mu,w}}(\mathbf{r}) +
\subsupi{E}{Hxc}{sr,\mu , {\it w}}[n_{\hat{\Gamma}^{\mu,w}}] ,
\end{eqnarray}
where the auxiliary ensemble density equals the physical one, 
\begin{eqnarray}\label{eq:ensemble_density}
n_{\hat{\Gamma}^{\mu,w}}(\mathbf{r})= (1-w)n_{\Psi_0^{\mu ,
w}}(\mathbf{r}) + w~n_{\Psi_1^{\mu ,
w}}(\mathbf{r})=n_{\hat{\Gamma}^{w}}(\mathbf{r}).
\end{eqnarray}
Using Equation~(\ref{eq:excitation_energy_exact}) leads to the following
exact expression for the first excitation energy~\cite{franck2014generalised,senjean2015linear},
\begin{eqnarray}\label{eq:excitation_energy_auxi+DD}
\omega & = & \mathcal{E}_1^{\mu , w} - \mathcal{E}_0^{\mu , w} + \left.
\dfrac{\partial \subsupi{E}{Hxc}{sr, \mu , {\it w}} [n]}{\partial w}
\right|_{n = n_{\hat{\Gamma}^{w}}} \nonumber \\
& = & \Delta \mathcal{E}^{\mu , w} + \Delta_{\rm xc}^{\mu , w},
\end{eqnarray}
where $\Delta \mathcal{E}^{\mu , w}$ is the auxiliary (weight-dependent)
excitation energy and $\Delta_{\rm xc}^{\mu , w}$ is 
the short-range exchange-correlation derivative discontinuity [the
short-range Hartree term does not depend on the weight as shown in
Equation~(\ref{eq:EHxc_decomposition_ensemble})]. As readily seen from
Equation~(\ref{eq:excitation_energy_auxi+DD}), the auxiliary excitation
energy is in principle not equal to the physical one. 

\subsection{Linear interpolation method}\label{subsec:LIM}

The construction of approximate ensemble exchange-correlation functionals is already a
challenge in the context of GOK-DFT~\cite{ParagiXCens,
Nagy_enseXpot,yang2014exact,pernal2015excitation}. In the case of range-separated ensemble DFT, an exact
adiabatic connection formula has been derived by Franck and
Fromager~\cite{franck2014generalised} but no approximations have been
developed so far. The simplest approximation consists in using the
(weights-independent) ground-state short-range
functional~\cite{PRA13_Pernal_srEDFT,senjean2015linear}, 
\begin{eqnarray}\label{eq:widfa}
\subsupi{E}{Hxc}{sr, \mu, {\mathbf{w}}} [n] \rightarrow \subsupi{E}{Hxc}{sr, \mu} [n] .
\end{eqnarray}
We shall refer to this approximation as {\it weights-independent
density-functional approximation} (WIDFA). The range-separated ensemble energy within WIDFA can be expressed as  
\begin{eqnarray}\label{rs-widfa-ens_energy}
\tilde{E}^{\mu,{\bf w}}&=&
\!\minover\limits_{\hat{\gamma}^{\mathbf{w}}} \!\left\{\!
{\rm Tr}\left[\hat{\gamma}^{\mathbf{w}}(\hat{T}+\hat{W}^{\rm
lr,\mu}_{\rm ee}+\hat{V}_{\rm ne})\right]
+ \!\! 
E_{\rm Hxc}^{\rm sr,\mu}[n_{\hat{\gamma}^{\mathbf{w}}}]
\! \right\}
\nonumber\\
&=&
{\rm Tr}\left[\hat{{\gamma}}^{\mu,\mathbf{w}}
(\hat{T}+\hat{W}^{\rm lr,\mu}_{\rm
ee}+\hat{V}_{\rm ne})
\right]
+E^{{\rm sr,\mu}}_{\rm Hxc}[n_{\hat{{\gamma}}^{\mu,\mathbf{w}}}],
\end{eqnarray}
where the minimizing ensemble density matrix $\hat{{\gamma}}^{\mu,\mathbf{w}}=
\sum_{k=0}^{M-1}w_k
\vert\tilde{\Psi}^{\mu,\mathbf{w}}_k\rangle\langle\tilde{\Psi}^{\mu,\mathbf{w}}_k\vert
$ fulfills the following set of self-consistent equations,
\begin{eqnarray}\label{eq:sc-srEDFT_eq_exact_general}
&&\Bigg(\hat{T}+\hat{W}^{\rm lr,\mu}_{\rm ee}+\hat{V}_{\rm ne}+
\int \ddroit\mathbf{r} 
\dfrac{\delta E_{\rm
Hxc}^{\rm sr,\mu}[n_{\hat{\gamma}^{\mu,\mathbf{w}}}]}{\delta
n(\mathbf{r})}
\hat{n}(\mathbf{r})
\Bigg)\vert \tilde{\Psi}^{\mu,\mathbf{w}}_k\rangle
\nonumber\\
&&=\tilde{\mathcal{E}}^{\mu,\mathbf{w}}_k\vert
\tilde{\Psi}^{\mu,\mathbf{w}}_k\rangle, \hspace{0.2cm} 0\leq k\leq M-1.
\end{eqnarray}
Note that the exact physical ensemble density matrix $\hat{\Gamma}^{\bf
w}$ and the exact
ensemble energy ${E}^{{\bf w}}$ are recovered from
Equation~(\ref{rs-widfa-ens_energy}) when $\mu\rightarrow+\infty$.\\

In the particular case of a two-state ensemble with weight $w$, the WIDFA excitation
energy defined as the derivative of the WIDFA ensemble energy with
respect to $w$ reduces to the approximate auxiliary excitation
energy~\cite{senjean2015linear}: 
\begin{eqnarray}\label{eq:excitation_energy_widfa}
\omega^{\mu , w}_{\rm WIDFA} = \dfrac{\ddroit \tilde{E}^{\mu , w}}{\ddroit w} = \tilde{\mathcal{E}}_1^{\mu , w} -\tilde{\mathcal{E}}_0^{\mu , w}  = \Delta \tilde{\mathcal{E}}^{\mu , w} .
\end{eqnarray}
As a result, it is both $\mu$- and $w$-dependent. The ambiguity in the
choice of the ensemble weight for computing excitation energies can be
overcome by means of linear interpolations between
equiensembles~\cite{senjean2015linear}. In other words, the exact
linearly-interpolated expression in Equation~(\ref{eq:excitation_energy_exact_ensemble}) is used rather
than the first-order-derivative-based expression in Equation~(\ref{eq:excitation_energy_exact}) in
order to compute approximate excitation energies. Both choices are of
course equivalent if exact wavefunctions and functionals are used. This
{\it linear interpolation method} (LIM) leads, for two states, to the
simple $\mu$-dependent (but $w$-independent) expression 
\begin{eqnarray}\label{eq:excitation_LIM}
\omega^\mu_{\rm LIM}  
 = 2(\tilde{E}^{\mu , 1/2} - \tilde{E}^{\mu , 0} ) .
\end{eqnarray}
Interestingly, LIM enables to define an effective derivative
discontinuity that exhibits, in helium for example, similar variations
in $w$ as the exact derivative discontinuity~\cite{senjean2015linear}:   
\begin{eqnarray}\label{eq:effective_DD}
\Delta_{\rm eff}^{\mu , w}=\omega^{\mu}_{\rm LIM}-\Delta
\tilde{\mathcal{E}}^{\mu , w}=\omega^{\mu}_{\rm LIM}- \omega^{\mu ,
w}_{\rm WIDFA}.
\end{eqnarray}
Note that $\Delta_{\rm eff}^{\mu , w}\rightarrow 0$ when $\mu \rightarrow
+\infty$. As a result, the extrapolation method of Savin~\cite{savin2014towards} cannot
be used in this context for obtaining more accurate variations of the
short-range derivative discontinuity with $w$ for a fixed and finite
value of $\mu$.\\

Let us finally mention that LIM can be easily extended to higher
excitations and degenerate states by considering the WIDFA ensemble
energy (denoted $\tilde{E}^{\mu,w}_{I}$ in the following) defined from
the following ensemble weights,  
\begin{eqnarray}
w_k= 
\begin{dcases}
    \dfrac{1-wg_I}{\displaystyle M_{I-1}} & 0\leq k\leq M_{I-1}-1,\\
    w              & M_{I-1} \leq k\leq M_I-1, 
\end{dcases}
\end{eqnarray}
where \begin{eqnarray}\label{eq:w_values_general_case}
0\leq w\leq\dfrac{1}{M_I},\nonumber\\
\displaystyle M_{I}=\sum^{I}_{L=0}g_L,
\end{eqnarray}
and $g_L$ is the degeneracy of the $L$th energy. The $I$th LIM excitation
energy can then be obtained as follows~\cite{senjean2015linear},
\begin{eqnarray}\label{eq:limXE_general_exp}
\omega_{{\rm LIM},I}^{\mu}&=&\dfrac{M_I}{g_I}\Big(\tilde{E}^{\mu,1/M_I}_{I}-\tilde{E}^{\mu,1/M_{I-1}}_{I-1}\Big)
+\dfrac{1}{M_{I-1}}\sum^{I-1}_{K=1}g_K\omega_{{\rm LIM},K}^{\mu}.
\end{eqnarray}
From the equality
$\tilde{E}^{\mu,1/M_{I-1}}_{I-1}=\tilde{E}^{\mu,0}_{I}$, it becomes
clear from Equation~(\ref{eq:limXE_general_exp}) that LIM interpolates the ensemble energy between equiensembles.
In the particular case of three non-degenerate states, the second
LIM excitation energy equals, with the notations of
Equation~(\ref{eq:excitation_LIM}), 
\begin{eqnarray}
\omega_{{\rm
LIM},2}^{\mu}=
3\Big(\tilde{E}^{\mu,1/3}_{2}-\tilde{E}^{\mu,1/2}\Big)+\dfrac{1}{2}\omega_{{\rm
LIM}}^{\mu}.
\end{eqnarray}

\subsection{Extrapolating excitation energies from range-separated ensemble energies}\label{subsec:LIM+Extrapolation}

In the spirit of Savin and coworkers~\cite{savin2014towards,
RebTouTeaHelSav-JCP-14,rebolini2015calculating}, we propose to extrapolate physical excitation energies
from the approximate LIM excitation energies. For that purpose, let us
introduce  $\eta=1/\mu$ and consider the $\mu\rightarrow+\infty$ limit
or, equivalently, $\eta\rightarrow0$. The WIDFA range-separated ensemble
energy in Equation~(\ref{rs-widfa-ens_energy}) can be
expanded through second order in $\eta$ as follows,
\begin{eqnarray}\label{widfa_ener_exp_largemu}
\tilde{{E}}^{1/\eta,\mathbf{w}}=E^{\mathbf{w}}+
\tilde{{E}}^{(-1),\mathbf{w}}\eta+\tilde{{E}}^{(-2),\mathbf{w}}\eta^2+\mathcal{O}(\eta^3),
\end{eqnarray}
where, according to the Hellmann--Feynman theorem,
\begin{eqnarray}\label{eq:gradient_ens_ener}
\tilde{{E}}^{(-1),\mathbf{w}}&=&
\left.\dfrac{\ddroit
\tilde{{E}}^{1/\eta,\mathbf{w}}}{\ddroit\eta}\right|_{\eta=0}
\nonumber\\
&=&
{\rm Tr}\left[\hat{\Gamma}^{\mathbf{w}}
\left.\dfrac{\partial \hat{W}^{\rm lr,1/\eta}_{\rm
ee}}{\partial\eta}\right|_{\eta=0}
\right]
+
\left.\dfrac{\partial E_{\rm Hxc}^{\rm
sr,1/\eta}[n_{\hat{\Gamma}^{\mathbf{w}}}]}{\partial \eta}
\right|_{\eta=0}.
\end{eqnarray} 
For large $\mu$ values, the short-range density-functional energy can be expanded as follows~\cite{srDFT,paolacorr},
\begin{eqnarray}
E^{{\rm sr,\mu}}_{\rm Hxc}[n]=
\dfrac{1}{\mu^2}E^{{\rm sr,(-2)}}_{\rm Hxc}[n]+\mathcal{O}\left(\dfrac{1}{\mu^3}\right).
\label{srfun_exp_largemu}
\end{eqnarray}
Therefore, the second term on the right-hand side of
Equation~(\ref{eq:gradient_ens_ener}) equals zero.
Moreover, according to Equation~(\ref{eq:w_ee_savin}),
\begin{eqnarray}
\dfrac{\partial w^{\rm lr,1/\eta}_{\rm ee}(r_{12})}{\partial
\eta}=-\dfrac{2}{\eta^2\sqrt{\pi}}e^{-r^2_{12}/\eta^2},
\end{eqnarray}
so that
\begin{eqnarray}\label{integral_ensem_intracule_dens}
{\rm Tr}\left[\hat{\Gamma}^{\mathbf{w}}
\dfrac{\partial \hat{W}^{\rm lr,1/\eta}_{\rm
ee}}{\partial\eta}
\right]
=-8\sqrt{\pi}\int_0^{+\infty}
\dfrac{r^2_{12}}{\eta^2}e^{-r^2_{12}/\eta^2}f_{\hat{\Gamma}^{\mathbf{w}}}(r_{12})\ddroit
r_{12},
\end{eqnarray}
where
$f_{\hat{\Gamma}^{\mathbf{w}}}(r_{12})=\sum_kw_kf_{\Psi_k}(r_{12})$ is
the exact physical ensemble intracule density (see, for example, Ref.~\cite{paolacorr}). It becomes clear from
Equation~(\ref{integral_ensem_intracule_dens}) that the first term on
the right-hand side of Equation~(\ref{eq:gradient_ens_ener}) is also
equal to zero, thus leading to $\tilde{{E}}^{(-1),\mathbf{w}}=0$.
In conclusion, for large $\mu$ values, the deviation of the WIDFA ensemble energy from the exact
one is at least of second order in $1/\mu$: 
\begin{eqnarray}
\tilde{E}^{\mu,{\bf w}}={E}^{{\bf w}}+\dfrac{1}{\mu^2}\tilde{E}^{(-2),{\bf w}}
+\mathcal{O}\left(\dfrac{1}{\mu^3}\right),
\end{eqnarray}
or, equivalently, 
\begin{eqnarray}\label{extrapol_ensener}
 {E}^{{\bf w}}=\tilde{E}^{\mu,{\bf w}}+\dfrac{\mu}{2}
\dfrac{\partial \tilde{E}^{\mu,{\bf w}}}{\partial \mu}
+\mathcal{O}\left(\dfrac{1}{\mu^3}\right).
\end{eqnarray}
A similar Taylor expansion can also be obtained for the auxiliary
energies,
\begin{eqnarray}\label{extrapol_auxiener_exact}
 E_k=\tilde{\mathcal{E}}_k^{\mu,{\bf w}}+\dfrac{\mu}{2}
\dfrac{\partial \tilde{\mathcal{E}}_k^{\mu,{\bf w}}}{\partial \mu}
+\mathcal{O}\left(\dfrac{1}{\mu^3}\right).
\end{eqnarray}
Thus we obtain from Equations~(\ref{extrapol_ensener}) and (\ref{extrapol_auxiener_exact}) the following extrapolated
expressions through second order in $1/\mu$ for the ensemble energy, 
\begin{eqnarray}\label{extrapol_ensener_second-order}
 {\tilde{E}}^{{\mu, \bf w}}_E =\tilde{E}^{\mu,{\bf w}}+\dfrac{\mu}{2}
\dfrac{\partial \tilde{E}^{\mu,{\bf w}}}{\partial \mu},
\end{eqnarray}
and the individual energies,
\begin{eqnarray}\label{extrapol_auxiener}
 \mathcal{\tilde{E}}^{{\mu, \bf w}}_{E,k} =\tilde{\mathcal{E}}_k^{\mu,{\bf w}}+\dfrac{\mu}{2}
\dfrac{\partial \tilde{\mathcal{E}}_k^{\mu,{\bf w}}}{\partial \mu},
\end{eqnarray}
respectively.
Note that Savin's extrapolation scheme (see
Equation~(\ref{extrapol_auxiener_exact_GSdens2})) is recovered from
Equation~(\ref{extrapol_auxiener}) in the ground-state density limit
($w_0=1$).\\

\begin{figure}
    \centering
\includegraphics[width=0.7\textwidth]{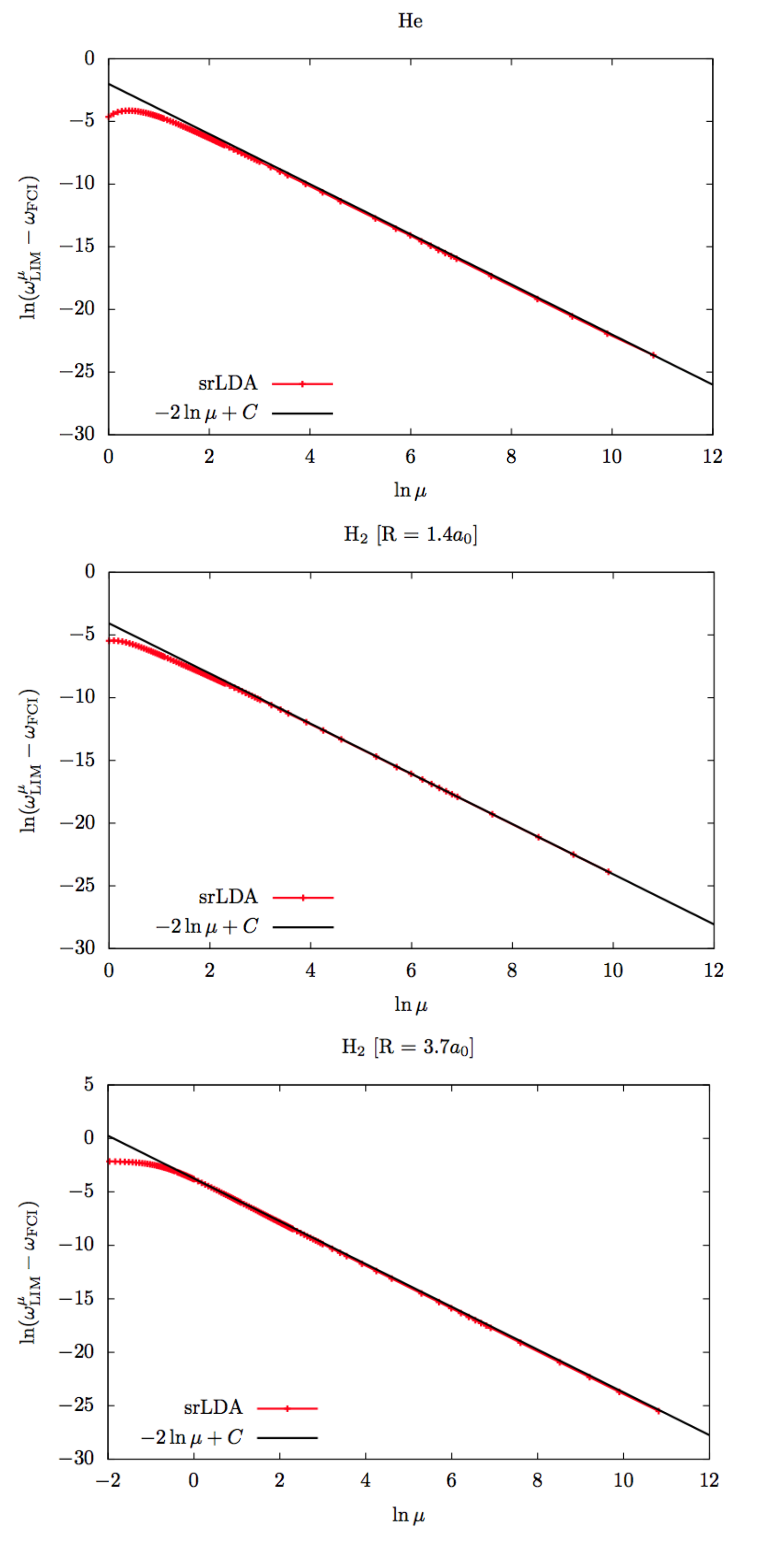}
    \caption{Graphical illustration of
Equation~(\ref{exp_LIMXE_largemu}) for two-state ensembles: $\{ 1^1S,
2^1S\}$ in He (top panel), $\{1^1\Sigma^+_g, 2^1\Sigma^+_g\}$ in H$_2$
at equilibrium (middle panel) and stretched (bottom panel) geometries. 
See text
for further details.}\label{fig:slope-2}
\end{figure}

Within LIM, the $I$th excitation energy $\omega^\mu_{{\rm LIM},I}$ is
simply expressed as the linear combination of range-separated
equiensemble energies that are all computed at the WIDFA level (see
equation~(\ref{eq:limXE_general_exp})). This expression becomes exact when $\mu\rightarrow+\infty$. Consequently, the exact $I$th excitation energy $\omega_I$ can be connected to the approximate LIM one as follows, 
\begin{eqnarray}\label{exp_LIMXE_largemu}
\omega^{\mu}_{{\rm LIM},I}=\omega_I +\dfrac{1}{\mu^2}\omega_{{\rm LIM}, I}^{(-2)}
+\mathcal{O}\left(\dfrac{1}{\mu^3}\right),
\end{eqnarray}
or, equivalently,
\begin{eqnarray}\label{extrapol_excener}
 \omega_I=\omega^\mu_{{\rm LIM},I}+\dfrac{\mu}{2}
\dfrac{\partial \omega^\mu_{{\rm LIM},I}}{\partial \mu}
+\mathcal{O}\left(\dfrac{1}{\mu^3}\right),
\end{eqnarray}
thus leading to the extrapolated LIM (ELIM) excitation energy expression
through second order in $1/\mu$, 
\begin{eqnarray}\label{extrapol_excener_2ndorder}
\omega^\mu_{{\rm ELIM},I} =\omega^\mu_{{\rm LIM},I}+\dfrac{\mu}{2}
\dfrac{\partial \omega^\mu_{{\rm LIM},I}}{\partial \mu}.
\end{eqnarray}
As illustrated in Fig.~\ref{fig:slope-2} for He and H$_2$, 
the LIM excitation energy does vary as $\mu^{-2}$ when $\mu$ is large,
thus showing that the second-order term in
Equation~(\ref{exp_LIMXE_largemu}) is not
zero.\\  

Let us finally mention that better extrapolations can be obtained by
considering higher-order derivatives in $\mu$~\cite{rebolini2015calculating}. This is left for future work.

\section{Summary}\label{Summary}

In this section we give a summary of all the approximate methods
introduced previously and whose performance is discussed in Sec.~\ref{sec:Results}.
All of them rely on the range separation of the two-electron repulsion
that is controlled by the parameter $\mu$ in the error function. A typical value is
$0.4$~\cite{rangeseparation2} but its influence on the results will be
investigated in details in the following. Note that, when $\mu=0$,
standard KS-DFT is recovered. On the other hand, wavefunction theory is
obtained when $\mu\rightarrow+\infty$. For intermediate values, the
long-range interaction is treated with wavefunction-based methods [full
configuration interaction (FCI) will be used in this work] and
short-range interactions are described by a complementary
$\mu$-dependent density
functional [a local functional will be used in this work].
Range-separated DFT can be extended to excited states by means of
ensembles. In the particular case of the first excitation, which is
mainly discussed in the paper, the ensemble consists of the ground- and
first-excited-state wavefunctions. A weight $w$, which can vary from 0
to 1/2, is assigned to the latter. The weight associated to the ground
state is $(1-w)$. The exact ensemble energy, which is
nothing but the weighted sum of the ground- and first-excited-state
energies, is linear in $w$. Therefore, its range-separated expression
should also be linear in $w$ (and $\mu$-independent) if the exact $w$-dependent short-range
ensemble functional were used. Note that, for a given $\mu$ value,
ground-state range-separated DFT is recovered when $w=0$. Returning to
the short-range functional for the ensemble, a simple approximation consists in employing
ground-state short-range functionals (a local one in our case) which has
therefore no weight dependence. As a result, the approximate range-separated
ensemble energy becomes both $\mu$- and $w$-dependent and, for a given
$\mu$ value, it exhibits curvature in $w$. Of course, this is not the
case in the exact theory so that the derivative of the ensemble energy
with respect to $w$ is $w$-independent and equal to the exact excitation
energy. The definition of approximate excitation
energies is then not trivial since the derivative of the range-separated
ensemble energy, which is referred to as auxiliary excitation
energy, varies with $w$. LIM is a way to remove this weight
dependence, simply by constructing, for a given $\mu$ value, the linear interpolation between the
ground state and the equiensemble ($w=1/2$). The slope of the
linearly-interpolated range-separated ensemble energy gives an
approximate excitation energy which is weight-independent by
construction. Of course, it still depends on $\mu$. When
$\mu\rightarrow+\infty$, both auxiliary and LIM excitation energies become exact. Therefore, the extrapolation
technique of Savin~\cite{savin2014towards} can be applied in this
context, thus leading for LIM to the extrapolated LIM (ELIM) approach. The
method can be generalized to an arbitrary number of states. An example will be given for
three states in H$_2$ along the bond breaking coordinate. 

\section{Computational details}\label{sec:computation}

Extrapolated auxiliary and LIM excitation energies (see
Equations~(\ref{extrapol_auxiener}) and
(\ref{extrapol_excener_2ndorder})) have been computed with a development
version of the DALTON program package \cite{DALTON2,DALTON}. Only the
spin-independent short-range local density approximation (srLDA)
\cite{savinbook,toulda} has been used. It was shown in a previous
work~\cite{senjean2015linear}
that the short-range Perdew-Burke-Ernzerhof-type functional of Goll~{\it
et al.}~\cite{Goll2005PCCP} gives rather similar results, at least
for the systems considered in this work. Basis sets are aug-cc-pVQZ
\cite{dunning1989gaussian, woon1994gaussian}. Orbitals relaxation and
long-range correlation effects have been treated self-consistently at
the FCI level. Calculations have been
performed on He and Be atoms as well as H$_2$ and HeH$^+$ molecules. For
Be, the 1$s$ orbital was kept inactive. The following two-state singlet
ensembles have been studied: $\{ 1^1S, 2^1S\}$ for He and Be,
$\{1^1\Sigma^+, 2^1\Sigma^+\}$ for the stretched HeH$^+$ molecule (R =
8.0 a$_0$), and $\{1^1\Sigma^+_g, 2^1\Sigma^+_g\}$ for H$_2$ at
equilibrium (R = 1.4 a$_0$) and stretched (R = 3.7 a$_0$) geometries. In
addition, the three-state singlet ensemble $\{1^1\Sigma^+_g, 2^1\Sigma^+_g,
3^1\Sigma_g^+\}$ in H$_2$ has been considered along the bond breaking
coordinate. In the latter case, comparison is made with
time-dependent multideterminant range-separated linear response
theory~\cite{fromager2013} using a FCI long-range interacting
wavefunction. Extrapolations have been obtained by finite
differences with $\Delta\mu=0.005 a_0^{-1}$. 

\section{Results and discussion}\label{sec:Results}

\subsection{Weight-dependence of the ensemble and auxiliary energies}

\begin{figure}
\centering
\resizebox{12cm}{!}{\includegraphics{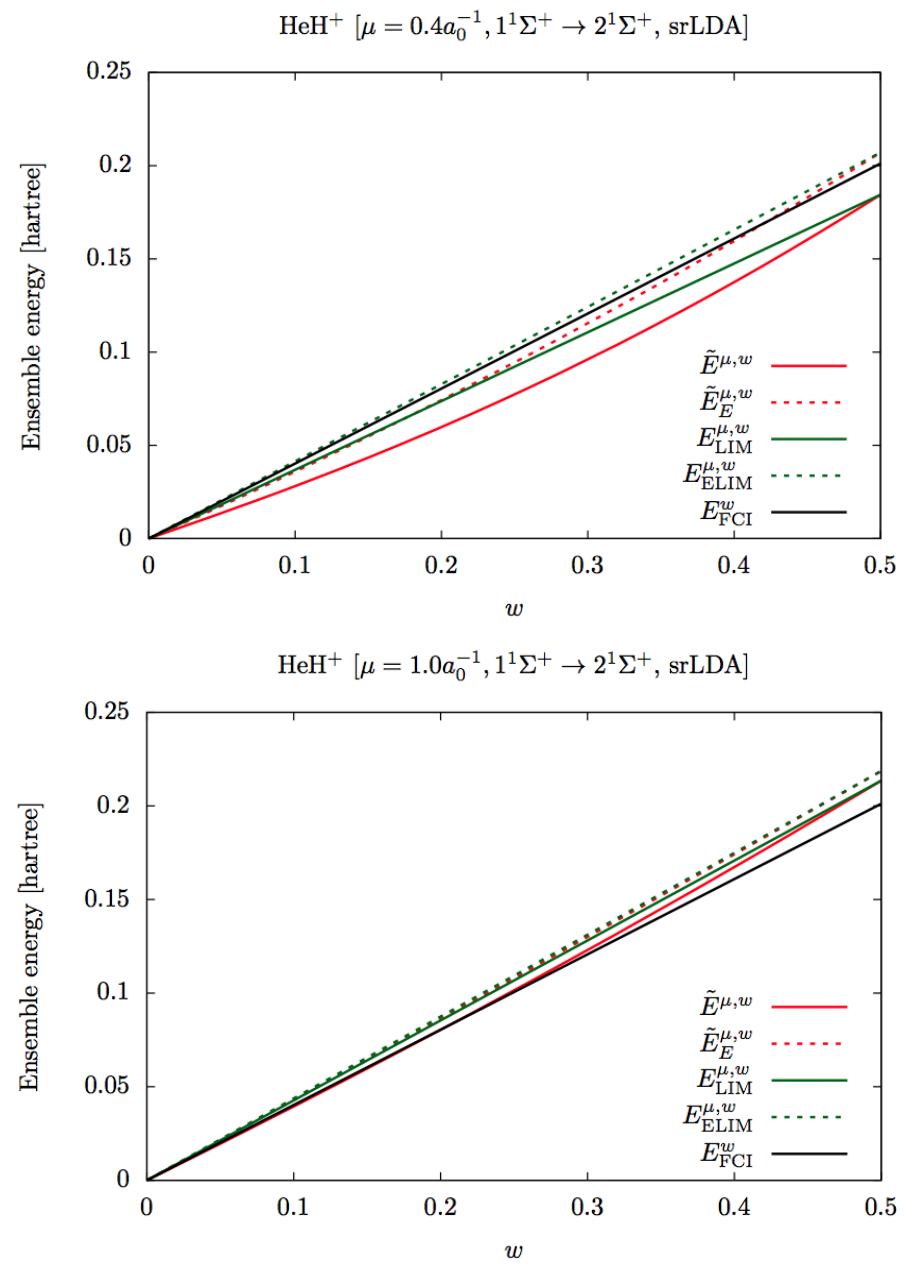}}
\caption{WIDFA and LIM ensemble energies 
computed 
with and without
extrapolation for the stretched HeH$^+$ molecule at $\mu = 0.4$ (top
panel) and $\mu = 1.0$ (bottom panel) when varying the ensemble weight
$w$. Energies at $w=0$ have been subtracted for ease of comparison. See
text for further details.}\label{fig:Ew_HeH+}
\end{figure}

The two-state $^1\Sigma^+$ WIDFA range-separated ensemble energy and its
extrapolation (see Equation~(\ref{extrapol_ensener_second-order})) have
been computed for the stretched HeH$^+$ molecule when varying the
ensemble weight $w$ for $\mu = 0.4$ and $\mu = 1$. Results are shown in
Fig.~(\ref{fig:Ew_HeH+}). Note that $2^1\Sigma^+$ is a charge-transfer
state. LIM and ELIM ensemble energies are also plotted. By construction,
both are linear in $w$, whereas the WIDFA ensemble energy is curved~\cite{senjean2015linear}.
When $\mu = 0.4$, the curvature is significantly reduced by the
extrapolation. In addition, the extrapolated WIDFA ensemble energy is much closer to
FCI. Note also that the slope at $w=0$, which corresponds to the
auxiliary excitation energy associated with the ground-state
density~\cite{senjean2015linear}, becomes very close to the FCI one after
extrapolation. This illustrates graphically the relevance of using  
extrapolated auxiliary energies for computing physical excitation
energies~\cite{RebTouTeaHelSav-JCP-14,rebolini2015calculating}. For $\mu = 1.0$, the extrapolation has less impact simply because
the WIDFA ensemble energy has a less pronounced curvature. The
contribution of the srLDA functional to the energy is simply reduced.   
ELIM and extrapolated WIDFA ensemble energies are almost
indistinguishable. This was expected since both WIDFA and LIM ensemble
energies (with or without extrapolation) become equal to the (linear)
FCI ensemble energy when $\mu\rightarrow+\infty$. Note, however, that
for $\mu = 1.0$ the extrapolation enlarges the
deviation of the ensemble energy from the FCI one. This will be analyzed
further in Sec.~\ref{sec:ELIM_XE_results}.\\ 

\begin{figure}
\centering
\includegraphics[width=0.7\textwidth]{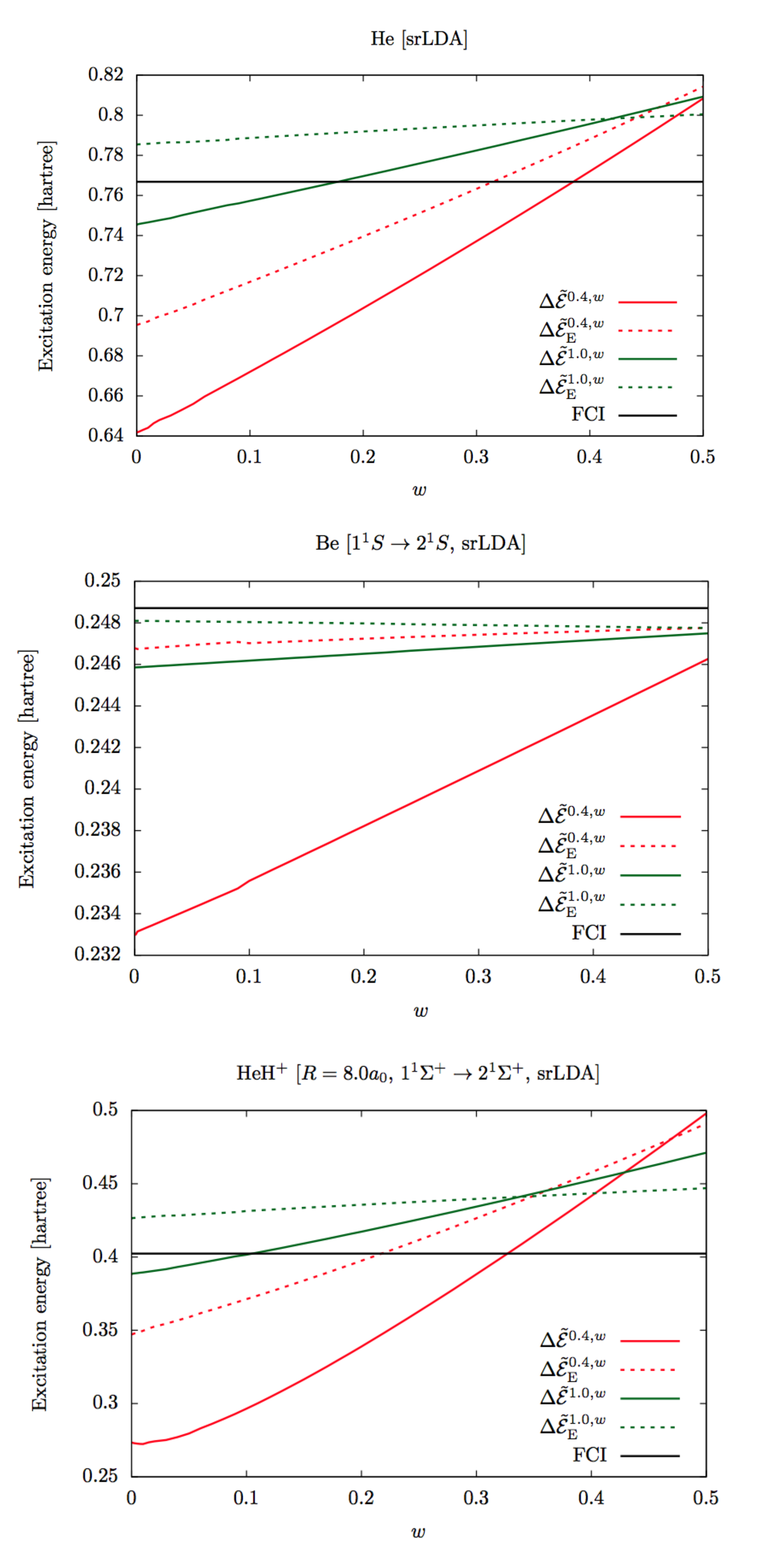}
\caption{Auxiliary excitation energies 
computed with and without
extrapolation for He (top
panel), Be (middle panel) and the streched HeH$^+$ molecule (bottom
panel) at  $\mu = 0.4$ and $\mu = 1.0$ when varying the ensemble weight
$w$. Comparison is made with FCI. See text for further details.}\label{fig:Auxi_He_Be_HeH+}
\end{figure}

\begin{figure}
\centering
\includegraphics[width=0.7\textwidth]{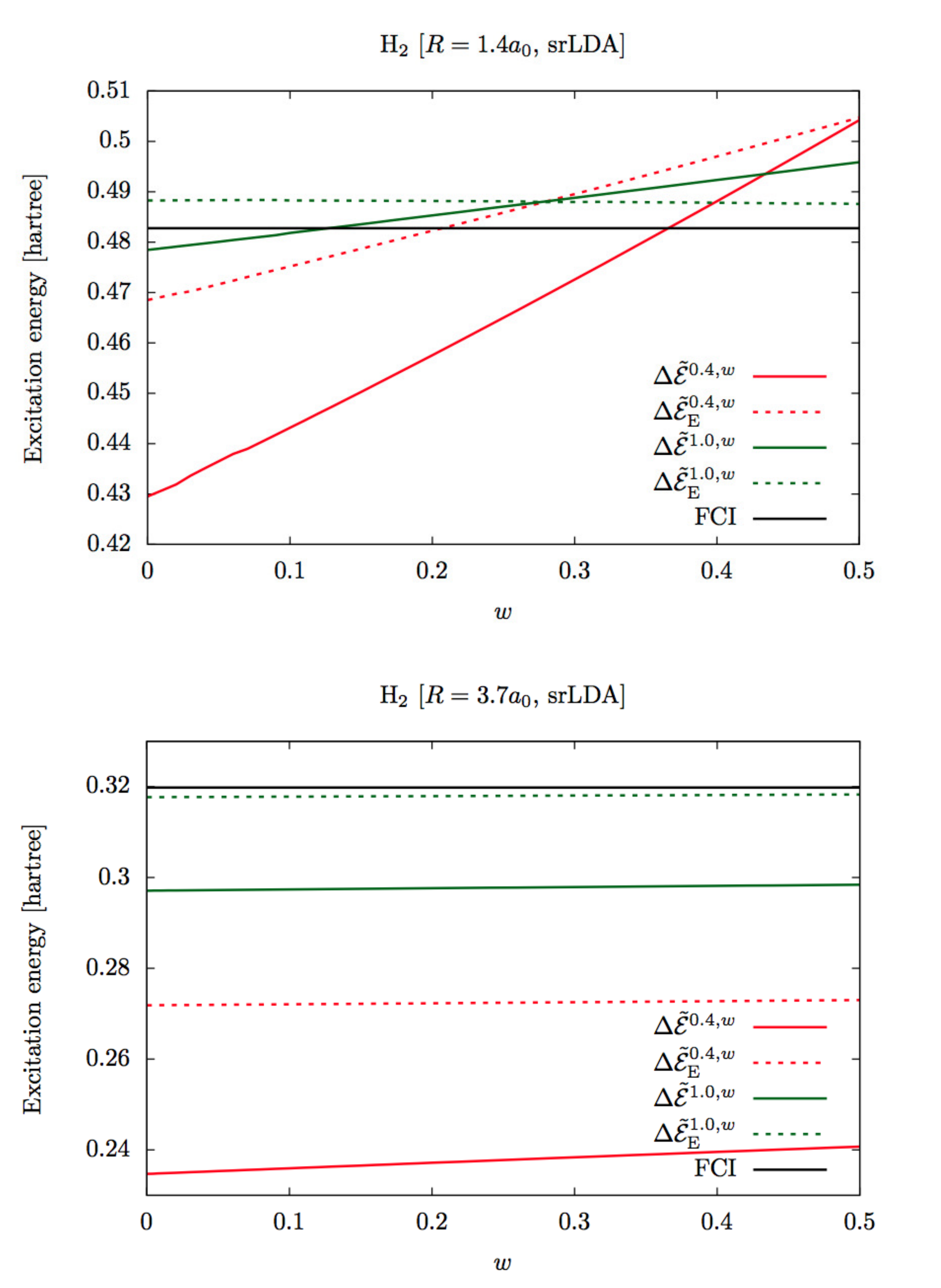}
\caption{First auxiliary excitation energy computed with and without
extrapolation for H$_2$ at equilibrium (top panel) and stretched (bottom
panel) geometries with $\mu = 0.4$ and $\mu = 1.0$ when varying the
ensemble weight
$w$. Comparison is made with FCI. See text for further details.}\label{fig:Auxi_H2}
\end{figure}

Let us now focus on the auxiliary excitation energies. In a previous
work~\cite{senjean2015linear}, some of the authors pointed out that the
latter can be strongly weight-dependent, thus motivating the formulation
of LIM. Extrapolation schemes are usually applied to auxiliary
excitation energies based on the ground-state
density~\cite{RebTouTeaHelSav-JCP-14,rebolini2015calculating}. In this section, we extrapolated
from weight-dependent auxiliary excitation energies (see
Equation~(\ref{extrapol_auxiener})), for analysis
purposes. Results are presented in Fig.~\ref{fig:Auxi_He_Be_HeH+} for
He, Be and HeH$^+$, and in Fig.~\ref{fig:Auxi_H2} for H$_2$ in
equilibrium and stretched geometries. 
As expected~\cite{senjean2015linear}, the weight-dependence of the
auxiliary excitation energy before extrapolation is reduced when
increasing $\mu$ from 0.4 to 1.0. For $\mu=0.4$ and 1.0, the extrapolation reduces the weight
dependence but, in general, it does not remove it completely.

\subsection{Extrapolated LIM excitation energies}\label{sec:ELIM_XE_results}

\begin{figure}
\centering
\includegraphics[width=0.7\textwidth]{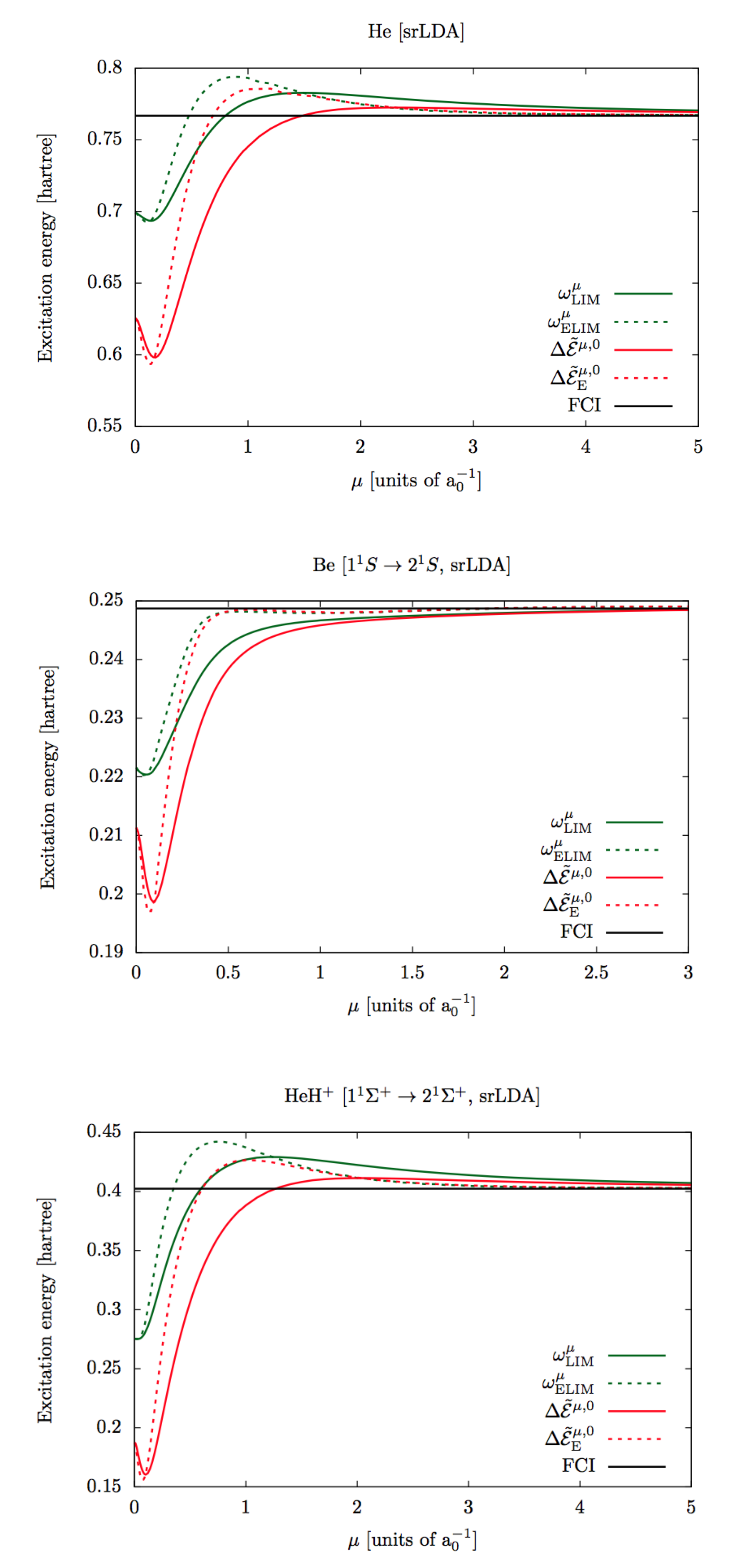}
\caption{LIM excitation energies computed with and without extrapolation
for He (top panel), Be (middle panel) and the stretched HeH$^+$ molecule
(bottom panel) when varying $\mu$. Comparison is made with the auxiliary
excitation energies and FCI.}\label{fig:LIM_He_Be_HeH+}
\end{figure}

\begin{figure}
\centering
\includegraphics[width=0.7\textwidth]{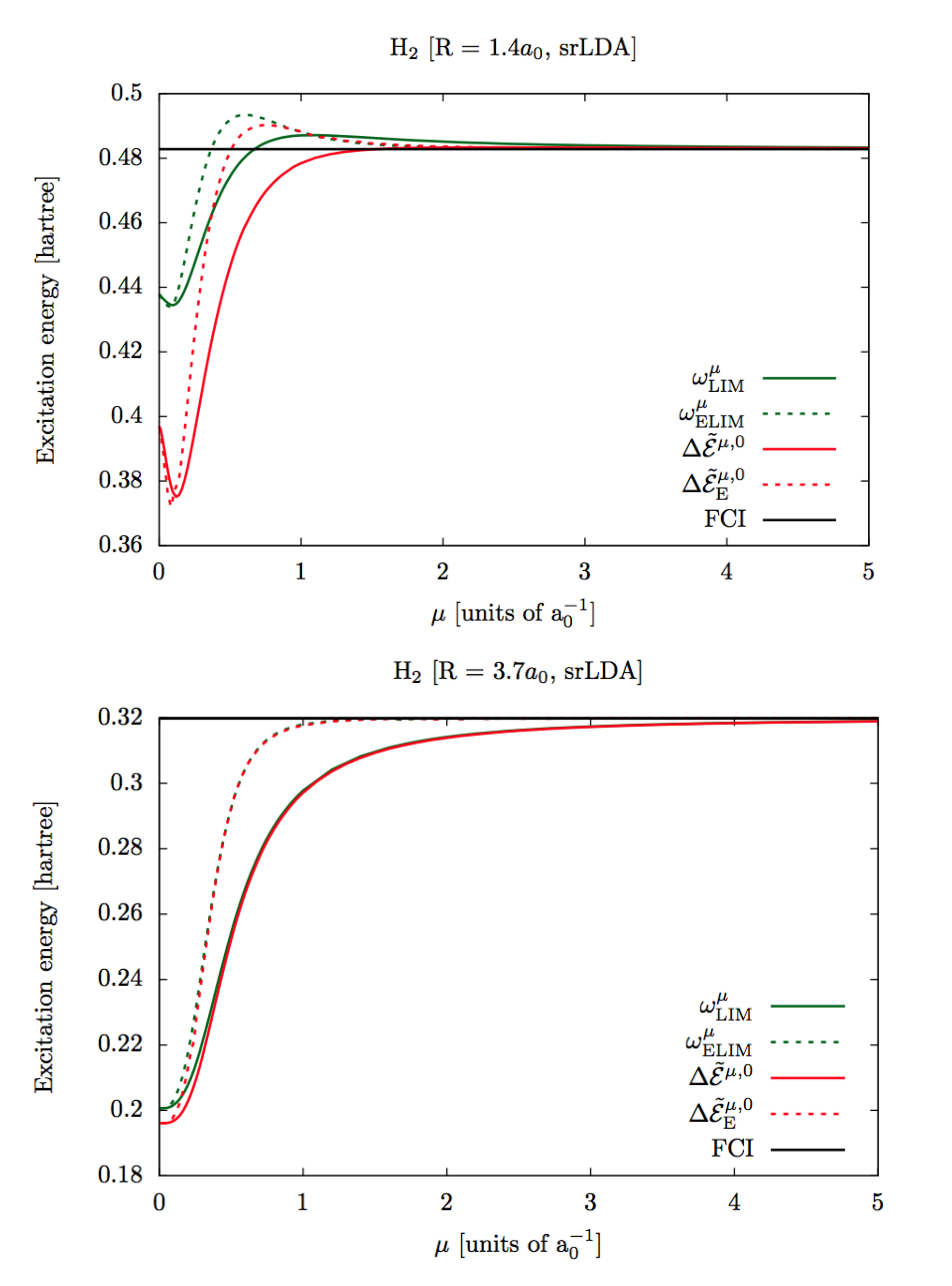}
\caption{First $^1\Sigma_g^+$ LIM excitation energy computed with and without
extrapolation for H$_2$ at equilibrium (top panel) and stretched (bottom
panel) geometries when varying $\mu$. Comparison is made with the auxiliary excitation
energy and FCI.}\label{fig:LIM_H2}
\end{figure}

As discussed previously, LIM provides approximate
excitation energies which are weight-independent by construction. In
this section, we investigate the impact of the extrapolation on the LIM
excitation energies when varying $\mu$. For analysis purposes,
comparison is made with the
extrapolated auxiliary
excitation energies based on the ground-state density. Results are presented in Figs.~\ref{fig:LIM_He_Be_HeH+}
and~\ref{fig:LIM_H2}.
Note that, by definition, the extrapolation has no effect at $\mu = 0$.
On the other hand, with or without extrapolation, FCI excitation
energies are recovered  when $\mu\rightarrow+\infty$. Without
extrapolation, LIM and auxiliary excitation energies are close to the
FCI results when $\mu \geq 2.0$ for Be and H$_2$ at equilibrium
(middle and bottom panels of Figs.~\ref{fig:LIM_He_Be_HeH+} and
\ref{fig:LIM_H2}, respectively), and when $\mu \geq 3.0$ for He,
HeH$^+$ and stretched H$_2$ (top and bottom panels of
Fig.~\ref{fig:LIM_He_Be_HeH+} and bottom panel of
Fig.~\ref{fig:LIM_H2}, respectively). We see that the extrapolation
improves the convergence in $\mu$ of both LIM and auxiliary excitation
energies towards the FCI values. However, better results are not
systematically obtained for all $\mu$ values. In He, HeH$^+$ and H$_2$
at equilibrium, for example, the extrapolation makes LIM deviate
from FCI when $0.5\leq\mu\leq1.0$. As shown 
in Ref.~\cite{rebolini2015calculating}, this is related to the
non-monotonic convergence of the excitation energies to the FCI
values. Such a deterioration of LIM is also observed when $\mu$ is close
to zero. The excitation energies are underestimated in that region and decrease
with $\mu$. This pattern was not observed for the auxiliary excitation
energies in
Refs.~\cite{rebolini2015calculating,RebTouTeaHelSav-JCP-14}, probably because the authors
computed accurate short-range potentials along the range-separated
adiabatic connection. It would be interesting, for rationalization
purposes, to compare exact and srLDA
LIM excitation energy 
expansions for small $\mu$ values. This is left for future work.\\  

Note finally that, for the typical $\mu = 0.4$ value~\cite{rangeseparation2},
ELIM gives relatively accurate results. In the particular case of the
stretched H$_2$ molecule, the improvement of the doubly-excited
$2^1\Sigma^+_g$ excitation energy after extrapolation is remarkable.  

\subsection{$2^1\Sigma^+_g$ and $3^1\Sigma^+_g$ excitation energies in H$_2$}\label{potential curves of H_2}

\begin{figure}
\centering
\includegraphics[width=0.7\textwidth]{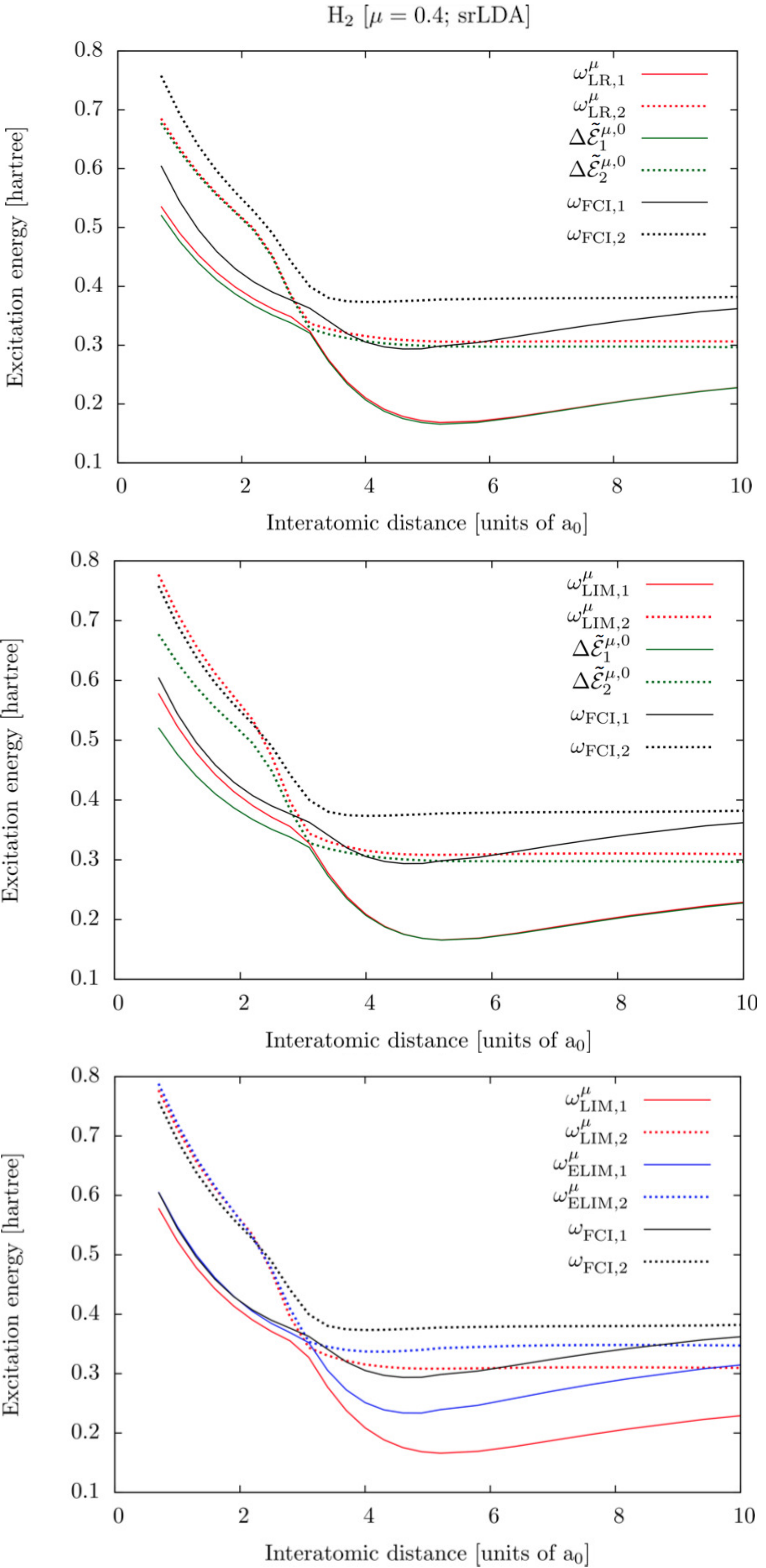}
\caption{First and second $^{1}\Sigma^+_g$ excitation energies computed
in H$_2$ along the bond breaking coordinate with $\mu = 0.4$. Top panel:
linear response (LR) versus auxiliary energies. Middle panel: LIM versus auxiliary
energies. Bottom panel: LIM versus ELIM. Comparison is made with FCI. 
See text for further details.
}\label{fig:H2_diss}
\end{figure}

$2^1\Sigma^+_g$ and $3^1\Sigma^+_g$ excitation energies in H$_2$ have been
computed along the bond breaking coordinate with various methods. The
range-separation parameter $\mu$ was set to the typical $\mu=0.4$
value~\cite{rangeseparation2}.
Results are presented in Fig.~\ref{fig:H2_diss}. For analysis
purposes, auxiliary excitation energies associated with the ground-state
density ($w_0=0$) are compared with FCI and time-dependent
multideterminant range-separated linear response (LR) results (see top panel). In the equilibrium region ($R\approx1.4 a_0$), LR
excitation energies are a bit closer to FCI values than the auxiliary
excitation energies. This is due to the short-range
kernel~\cite{fromager2013,senjean2015linear}. For larger
bond distances ($3 a_0 \leq R \leq 3.5 a_0$), the avoided crossing
obtained at the FCI level is not well reproduced. Auxiliary energies
give almost a crossing while LR slightly increases the gap between the
two excitation energies, once again, because of the short-range kernel.
When approaching the dissociation limit, the latter does not contribute
anymore to the $2^1\Sigma^+_g$ LR excitation energy that becomes
identical to the auxiliary one. This is due to the doubly-excited
character of the excitation which does not induce changes in the density
through first order~\cite{fromager2013,senjean2015linear}. Note the slight difference between
$3^1\Sigma^+_g$ auxiliary and LR excitation energies. In this case, the
short-range kernel does contribute.
Let us finally point out that both
range-separated approaches underestimate the excitation energies.\\           

Let us now discuss the performance of LIM (middle panel of
Fig.~\ref{fig:H2_diss}). It is
remarkable that, in the equilibrium region, LIM excitation energies are
much more accurate than the auxiliary ones for both $2^1\Sigma^+_g$ and
$3^1\Sigma^+_g$ states. The avoided crossing is relatively well located
within LIM but, like at the LR level, it is not well reproduced (the states are too close in
energy). For larger bond distances, $2^1\Sigma^+_g$ LIM and auxiliary excitation
energies become identical, as expected~\cite{senjean2015linear}. On the other hand, those
differ slightly for the $3^1\Sigma^+_g$ state. In summary, LIM
significantly improves on
both LR and auxiliary excitation energies only in the equilibrium
region.\\  

Turning to ELIM results (bottom panel of Fig.~\ref{fig:H2_diss}), we first notice that, in the
equilibrium region, the $2^1\Sigma^+_g$ excitation energy curve is
almost on top of the FCI one, as expected from the top panel of
Fig.~\ref{fig:LIM_H2}. The extrapolation has less impact on the
$3^1\Sigma^+_g$ state. For the latter, it actually enlarges the
deviation from FCI. This is simply due to the fact that, without
extrapolation, LIM already overestimates the $3^1\Sigma^+_g$ excitation
energy. The positive slope in $\mu$ of the LIM excitation energy in the
vicinity of $\mu=0.4$, exactly like for the $2^1\Sigma^+_g$ state (see
top panel in Fig.~\ref{fig:LIM_H2}), increases further the excitation
energy.  When the bond is stretched, ELIM improves on both
individual excitation energies (both become much closer to the reference FCI
values). The effect of the extrapolation becomes significant when
approaching the dissociation limit. However, even though the avoided
crossing remains relatively well located within ELIM, the two states
become closer in energy after extrapolation. This shows how challenging
it is to develop a reliable multideterminant state-averaged
range-separated DFT for modelling (avoided) crossings. It is still
unclear if a ghost interaction correction would provide better results.
Work is currently in progress in this direction.       

\section{Perspective: A remedy for the {ghost interaction} error in
multideterminant range-separated DFT}\label{sec:Perspective}

Unlike in Hartree-Fock theory, the exact cancellation of Hartree and
exchange terms, for one electron systems, is not possible in KS-DFT when using
local or semi-local functionals, thus inducing so-called
self-interaction errors. However, the use of orbital-dependent
functionals can correct for this. Another type of error, referred to as
ghost interaction error~\cite{ensemble_ghost_interaction}, is introduced
in ensemble DFT calculations when an ensemble density is inserted into
the usual (or short-range) Hartree density functional. The ghost
interaction terms contain products of densities associated with
different states. The use of local or semi-local exchange-correlation
functionals cannot be expected to correct for such
errors~\cite{pastorczak2014ensemble}. Note that, in multideterminant
range-separated ensemble DFT, densities are calculated from
multideterminant wavefunctions so that standard ghost interaction
correction schemes, which have been developed in the context of GOK-DFT, cannot
be used straightforwardly.\\  

In the present section we propose an alternative decomposition of the
short-range ensemble exchange-correlation energy which has the
advantage of being ghost-interaction-free.  For that purpose we use the
concept of multideterminantal (md) short-range exact exchange introduced by
Toulouse~{\it et al.}~\cite{Toulouse2005TCA} in the context of
ground-state range-separated DFT and extend it to ensembles as follows,
\begin{eqnarray}\label{srHxc_alt_decomposition}
E_{\rm{Hxc}}^{\rm{sr,\mu,\mathbf{w}}}\left[n\right] =
E_{\rm{H}}^{\rm{sr,\mu}}\left[n\right] +
E_{\rm{x,md}}^{\rm{sr,\mu,\mathbf{w}}}\left[n\right] +
E_{\rm{c,md}}^{\rm{sr,\mu,\mathbf{w}}}\left[n\right],
\end{eqnarray}
where
\begin{eqnarray}\label{md_exchange}
E_{\rm{x,md}}^{\rm{sr,\mu,\mathbf{w}}}\left[n\right] = \text{Tr}\left[\hat{\Gamma}^{\rm{\mu,\mathbf{w}}}[n]\hat{{W}}_{\rm{ee}}^{\rm{sr,\mu}}\right] - E_{\rm{H}}^{\rm{sr,\mu}}[n],
\end{eqnarray}
and, according to Equations~(\ref{eq:EHxc_decomposition_ensemble}) and
(\ref{eq:exc_functional_ensemble}), 
\begin{eqnarray}\label{md_correlation}
E_{\rm{c,md}}^{\rm{sr,\mu,\mathbf{w}}}[n] =
E_{\rm c}^{\rm{sr,\mu,\mathbf{w}}}[n] + \text{Tr}\left[
\hat{\Gamma}_{{s}}^{\rm{\mathbf{w}}}[n]\hat{{W}}_{\rm{ee}}^{\rm{sr,\mu}}\right]
- \text{Tr}\left[
\hat{\Gamma}^{\rm{\mu,\mathbf{w}}}[n]\hat{{W}}_{\rm{ee}}^{\rm{sr,\mu}}
\right].
\end{eqnarray}
Note that the expression
in Equation~(\ref{md_exchange}) involves the
ensemble density matrix of the long-range interacting system with
ensemble density $n$ rather than the noninteracting one (see
Equation~(\ref{eq:exc_functional_ensemble})
for comparison). Finally, the
complementary short-range correlation functional $E_{\rm{c,md}}^{\rm{sr,\mu}}[n]$
that is adapted to the exact multideterminant short-range exchange of
Toulouse~{\it et al.}~\cite{Toulouse2005TCA} is recovered in the
ground-state limit ($w_0=1$). 
Since, according to Equation~(\ref{eq:self_consistent_savin_ensemble}),
\begin{eqnarray}
\hat{\Gamma}^{\rm{\mu,\mathbf{w}}}[n_{\hat{\Gamma}^{\mu,
\mathbf{w}}}]=\hat{\Gamma}^{\mu, \mathbf{w}},
\end{eqnarray}
 combining the alternative decomposition in
Equation~(\ref{srHxc_alt_decomposition}) with
Equation~(\ref{eq:ensemble_energy_savin_ensemble}) leads to the exact ensemble energy expression: 
\begin{eqnarray}\label{ensemble_energy_altsep}
E^{\rm{\mathbf{w}}}= 
\text{Tr} \left[ \hat{\Gamma}^{\rm{\mu,\mathbf{w}}}\left( \hat{{T}} +
\hat{{W}}_{\rm ee} + \hat{{V}}_{\rm ne} \right) \right] +
E_{\rm{c,md}}^{\rm{sr,\mu,\mathbf{w}}}\left[n_{\hat{\Gamma}^{\rm{\mu,\mathbf{w}}}}\right].
\end{eqnarray}
Like in the ground-state theory, this expression {\it cannot} be
variational with respect to the ensemble density matrix, otherwise
double counting problems would occur~\cite{manusroep2013}. Optimized effective
potential techniques should be applied in this context to avoid such
problems~\cite{manusroep2013}. A simple approximation would consist in
replacing the
exact auxiliary ensemble density matrix with
the one obtained at the WIDFA level 
(see Equation~(\ref{rs-widfa-ens_energy})),
\begin{eqnarray}
\hat{\Gamma}^{\rm{\mu,\mathbf{w}}}\rightarrow\hat{\gamma}^{\rm{\mu,\mathbf{w}}}.
\end{eqnarray} 
Moreover, in the spirit of WIDFA, one could use the ground-state 
correlation functional, for which a local density approximation has been
developed~\cite{Paziani2006PRB}, rather than the ensemble functional
(for which no approximations have been developed so far),  
\begin{eqnarray}
E_{\rm{c,md}}^{\rm{sr,\mu,\mathbf{w}}}[n]\rightarrow
E_{\rm{c,md}}^{\rm{sr,\mu}}[n].
\end{eqnarray}
Work is currently in progress in this direction. Note that the resulting
approximate ensemble energy will be ghost interaction-free. 
Interestingly, the alternative separation of short-range ensemble exchange and correlation
energies is a way to introduce (implicitly) weight dependence into the
short-range ensemble functional. The resulting approximate ensemble
energy may also have less curvature in the ensemble weights than the
WIDFA one. This should be investigated numerically. Note finally that
LIM can also
be applied in this context in order to compute excitation energies. 

\section{Conclusion}\label{sec:Conclusion}

The extrapolation scheme initially proposed by
Savin~\cite{savin2014towards} in the context of
ground-state range-separated DFT has been extended
to ensembles of ground and excited states. This can be achieved when
expanding the range-separated ensemble energy in powers of $1/\mu$,
where $\mu$ is the parameter that controls the range separation of the
two-electron repulsion. Combining this approach with the recently
proposed linear interpolation method (LIM)~\cite{senjean2015linear} enables to compute excitation
energies that, by construction, do not depend on the choice of the
ensemble weights. We have shown on a small test set consisting of He,
Be, H$_2$ and HeH$^+$ that, for the typical $\mu=0.4$ value, the
extrapolated LIM (ELIM) can provide accurate (sometimes very accurate) excitation energies
even for charge-transfer and doubly-excited states. It was also shown
that, in the stretched H$_2$ molecule, the extrapolation can improve on
excitation energies individually but the relative excitation energy can
be deteriorated, thus leading to an inaccurate description of avoided
crossings. Such problems should be investigated further in the future in
order to turn multi-determinant range-separated DFT into a reliable
computational tool for modeling photochemistry, for example.
A potential source of errors in LIM and ELIM calculations is the
so-called ghost-interaction error that is introduced when inserting an
ensemble density into the short-range Hartree functional. We proposed an
alternative separation of ensemble short-range exchange and correlation
energies which, in principle, enables the calculation of
ghost-interaction-free excitation energies in the context of
multideterminant range-separated ensemble DFT. Work is currently in
progress in this direction.

\section*{Acknowledgments}

The authors are pleased to dedicate this work to Andreas Savin on the
occasion of his 65th birthday (congratulations Andreas and thanks for
everything !).
E.~F. would like to thank Julien Toulouse for stimulating discussions on
extrapolation schemes. E.~F. finally acknowledges financial support from the LABEX "Chemistry of complex systems"
and the ANR (MCFUNEX project). E.~D.~H. acknowledges the Villum Kann
Rasmussen foundation for a post-doc fellowship.


\label{lastpage}

\bibliographystyle{tMPH}


\newcommand{\Aa}[0]{Aa}

\end{document}